\documentclass[structabstract]{aa}
\usepackage{graphicx}
\usepackage{lscape}
\usepackage{amsmath}
\usepackage{txfonts}
\usepackage{url}
\usepackage{natbib}
\usepackage{xcolor}
\begin{document} 

\title{Evolution of single B-type stars\\ with a large angular momentum content}
\titlerunning{B-type stars with a large angular momentum content}
\author{Anah\'i Granada \& Lionel Haemmerl\'e}
\institute{Geneva Observatory, University of Geneva, Maillettes 51, CH-1290 Sauverny, Switzerland\\}

\date{Received ; accepted } 
\abstract
{The Geneva Stellar Evolution Group has recently presented an extended database of rotating stellar
models at three different metallicities for nine different initial
rotation parameters and ten different masses corresponding to spectral types from early- F to late- O. 
With these grids we have contributed to the understanding of the evolution of single rotating stars, and we intend to 
use them to produce synthetic stellar populations that fully account for the effects of
stellar rotation. However, up to now we still lacked stellar evolutionary tracks that rotate close to the critical limit during the whole main-sequence (MS) phase.
This occurs because the flat internal profile of rotation imposed at the zero-age main sequence (ZAMS) 
is modified by the action of meridional currents immediately after the ZAMS, causing the surface rotational velocity to decrease abruptly until it reaches a quasi-stationary state.}
{We compute stellar models with non-solid rotation at the ZAMS, which allows us to
 obtain stellar evolutionary tracks with a larger content of angular momentum that rotate close to the breakup limit throughout the whole MS.}
{We produced stellar models by removing the assumption that stars rotate as solid bodies at the ZAMS. We obtained the stellar structure 
at the ZAMS with a differentially rotating profile for three different metallicities by performing pre-MS calculations and by proposing ad hoc initial rotational profiles. 
We then computed the MS evolution and later phases of stellar evolution of these models, which attain rotational equatorial velocities close 
to the critical limit throughout their whole MS phase.}
{Stellar models with solid rotation at the ZAMS adequately represent the overall characteristics and evolution of 
differentially rotating models of identical angular momentum content, but with a 
lower initial surface rotational velocity rate, at Z=0.014, Z=0.006, and Z=0.002.   
For models with solid rotation at the ZAMS we therefore recommend to use as the initial rotational rate the values derived once the quasi-stationary 
state is reached, that is, after the abrupt decrease in surface velocity.  
By producing stellar structures at the ZAMS with differentially rotating profiles and larger angular momentum content than in our previous works,   
we obtain models that rotate close to the critical limit throughout the whole main sequence.
These models have a longer MS lifetime and a higher surface chemical enrichment already at the end of the MS, particularly at Z=0.002.
Interestingly, the initial equatorial rotational velocities are virtually metallicity independent for all stellar models we computed in the B-type star range with the same mass and angular momentum content at the ZAMS. 
If, as some observational evidence indicates, B-type stars at Z=0.002 rotate with a higher equatorial velocity at the ZAMS than stars with Z=0.014, 
our finding would indicate that the angular momentum content of B-type stars in the SMC is higher than their Galactic counterparts.}
   {}


\maketitle
\section{Introduction}
Rotation has proved to play a relevant role on the formation, 
evolution, and the ultimate fate of stars \citep[see e.g.][]{Maeder2009a} and, as a consequence, 
it has deep implications for the behaviour of different populations of stars in clusters and galaxies. 
 Therefore, it is important to study rotating stellar populations to gain a better understanding  of the physics of rotation in stars.
To do this, it is necessary to explore, both from a theoretical and an observational point of view, 
groups of stars in which the effects of rotation are strong. 
 In particular, the domain of B-type stars has been known for a long time to offer an exceptional laboratory for studying stellar 
rotation \citep{Slettebak1966,Sackmann1970}[e.g.], because it hosts the largest portion of rapidly 
rotating stars. Among them, we find Be stars \citep[see][for a recent review on these rapidly rotating objects]{Rivinius2013}.

With the aim of studying in detail the effects of rotation troughout the B-type stellar range, \citet{Georgy2013a} 
 presented an extended database of models of rotating stars at three different metallicities for nine different initial
rotation parameters and ten different masses (between 1.7 and 15 M$_{\sun}$) at the zero-age main sequence (ZAMS). 
This was achieved by the numerous improvements that were implemented in the Geneva stellar evolution code \citep{Ekstrom2012a}, 
which enable computing models of rotating stars, even those that rotate at the critical limit and experience episodes of mechanical mass loss 
\citep{Georgy2013a,Granada2013a}. 

{It is important to to keep in mind that in order to do this, many assumptions and simplifications were made,  
that enabled us to account for physical processes that are not unidimensional, such as rotation itself, convection, turbulent mixing, stellar winds, etc. 
Many of these processes are still poorly understood and remain treated with simple parametrized approximations.
That is the case, for instance, of the choice of an overshooting parameter in convective regions, calibrated to ensure an 
adequate reproduction of the width of the MS sequence band in the Hertzprung-Russell diagram (HRD), or the prescription used for the diffusion coefficients 
that express the transport of angular momentum in rotating models\footnote{ An interesting discussion of the impact of 
different prescriptions for the diffusion coefficients included in the shellular rotating models was provided by \citet[][]{Meynet2013}.}, among others.

To improve the prescriptions we can benefit from 2D/3D models, which provide a more realistic description of the structure of rotating stars.
For instance, \citet{EspinosaLara2011} have succedeed in obtaining a better description of the latitude dependency of the temperature in rapidly rotating stars. 
Other authors \citep[e.g.][]{Deupree2012} have calculated frequency spectra for rotating pulsators, which is 
useful to interpret the asteroseismic data. These types of data are most promising to set contraints on the internal structure of the star 
and its rotational properties and evolution \citep{Eggenberger2012,Neiner2012}. 
Multidimensional hydrodynamical simulations can als provide constraints on the treatment of turbulence \citep{Arnett2011} and 
hence better describe the transport of angular momentum inside the star. Until many of the new constraints prove 
to be better than the existing ones, we continue to use the same prescriptions that 
have been used in previous computations with the Geneva stellar evolution code. 

We recall some of the main assumptions made for computing grids of rotating models.
\begin{itemize}
 \item Shellular rotation hypothesis: $\Omega$ is constant on isobars, because of strong horizontal turbulence \citep{Zahn1992a}. 
Under this assumption, there is no latitude dependency of stellar rotation.   
 \item The shape of a rotating star is given by the Roche model, which assumes a central condensation of the stellar mass. 
In this framework, the surface of the star only depends on $\omega=\Omega/\Omega_{\rm crit}$, 
with $\Omega$ the angular velocity of the stellar surface and $\Omega_{\rm crit}$ the breakup angular velocity. 
Even for stars with a large angular momentum content such as those we present here the stellar interior rotates well below its critical limit,
while the surface of the star is rotationally distorted, and 1D models remain an acceptable approximation.
 \item Because of the Von Zeipel effect, the stellar winds are anisotropic. In our computations we used the prescripions described by \citet{Georgy2011a}.
 \item Throughout the evolution of a rapidly rotating star, the equatorial rotational velocity can reach the break-up limit. 
In our models, we assume that stars reaching the critical limit will undergo mechanical mass and angular momentum losses, which we model as
removing the overcritical layers of the star. However, the details of mechanical mass removal close to the critical limit are still not understood,
 and our models also lack the physical process to effectively push the matter outwards.
\end{itemize}}
We refer to \citet{Ekstrom2008b,Ekstrom2012a} and \citet{Georgy2013a,Georgy2011a} for extensive explanations and details of the physical processes and assumptions
made in our calculations.

We plan to use our rotating grids to produce synthetic populations of stars and study their evolution in time, fully 
accounting for mass, rotation and metallicity distributions, with our new population synthesis code \citep{Georgy2014a}. 
The stellar models of the database were computed assuming solid-body rotation at the ZAMS, 
with the underlying assumption that pre- main sequence (pre-MS) stars are fully
convective, similarly to the work by \citet{Heger2000a}, and assuming that convection drives rigid-body rotation\footnote{As mentioned above, it is beyond of the scope of this article to explore other rotation prescriptions on the convective regions, 
but this would certainly have an impact on the amount of angular momentum a star can have in its interior for a certain surface angular velocity, and therefore on our results.}.
In our models, the action of meridional currents modifies the flat internal rotation profile 
just after the ZAMS when the solid-body constraint is released, which leads to a strong decrease of the surface rotation.
In this scenario, our models therefore begin their evolution with a slow-down of their surface
rotational velocity. If the initial velocity is high enough, they can reach the critical limit after
some evolution on the main sequence (MS). As a consequence, the initial conditions
of the calculations prevent obtaining very
rapidly rotating stars at very young ages, even when imposing
a high value of $\omega_{ini}$. Because we intend to study the evolution of rapidly rotating stellar populations,
it is certainly a problem that we are unable
to produce rapidly rotating stars from the ZAMS because of the strong assumption of an initial flat rotating profile.

The recent work by \citet{Haemmerle2013}, who computed models of pre-MS stellar evolution, can help us solve this problem.
On the one hand, it allows us to study the validity of the assumption of solid-body rotation at the ZAMS, that is, to understand how stars
with the same mass, metallicity and angular momentum content but different internal rotation profile at the ZAMS evolve.
On the other hand, 
it permits us to produce very rapidly rotating stellar models from the ZAMS and throughout the whole MS lifetime. 

Throughout the present article we extend and complement the work by \cite{Georgy2013a}.
In Section 2 we describe the assumptions made by \citet{Haemmerle2013} to obtain the internal rotation profile at the ZAMS, 
and describe the results obtained by these authors, which are relevant to the present work. Because we 
present here models with different metallicities, we also show that the results obtained for Z=0.014 are valid for lower metallicities as well. 
In Section 3 we present our new models of rapidly rotating stars from the ZAMS with a large angular momentum content, we describe their main characteristics,
 and show how they compare with the grids presented by \citet{Georgy2013a}. We conclude in Section 4.

\section{Relevance of the internal rotational profile of a star at the ZAMS for its subsequent evolution}

\citet{Haemmerle2013} showed that stars that form through different pre-MS accretion 
prescriptions, but have the same content of total angular momentum at the ZAMS, follow the same subsequent evolution in terms
of evolutionary tracks, surface velocities, and abundances, independently of their formation history.
These authors computed pre-MS tracks for stars between 2 and 22 M$_{\sun}$ at solar metallicity,
using the Geneva stellar evolution code \citep{Eggenberger2008,Ekstrom2012a},
including shellular rotation, and following different scenarios, with accretion or at constant mass.
The initial model, at the top of the Hayashi line\footnote{
In the accretion scenario, the initial model is a 0.7 M$_{\sun}$ hydrostatic core
at the top of the Hayashi line, which corresponds to this mass value.}, is fully convective, and solid-body rotation is assumed in convective 
zones.
During pre-MS evolution, large regions of the star become progressively radiative and a gradient of internal angular rotation velocity ($\Omega$) develops,
so that the rotation profile obtained when the star reaches the ZAMS is not flat.
For models reaching $\rm v_{surf}/v_{crit}=0.4$ on the ZAMS in the accretion scenario,
the ratio between the central and the surface value of $\Omega$ is 1.8 for 2 M$_{\sun}$, and 1.2 for 14 M$_{\sun}$.
Similar values were obtained for constant mass.
Then, \citet{Haemmerle2013} were able to compare the MS evolution of stars with this rotation profile on the ZAMS,
with the MS evolution of stars rotating as a solid body on the ZAMS. The authors obtained that
after a re-adjustment phase that lasts for $\sim15\%$ of the total MS time, models with the same mass and angular momentum content with
 different internal rotation profiles at the ZAMS converge to the same internal rotation profile. 
Or, in other words, they found that the rotation profile after this re-adjustment phase is only determined by the total angular momentum of the star, and
that the evolutionary tracks and the surface abundances show no significant differences throughout later evolutionary phases.
As a consequence, for a given angular momentum content, the choice of the rotation profile on the ZAMS
has no impact on the subsequent evolution, except in the very beginning of the MS.

This result allows circumventing the limitation in rotation velocity for models with solid-body rotation on the ZAMS, which unavoidably undergo a
sharp decrease in the surface rotational velocity. We can now produce models with a non-flat internal rotation profile  
with a larger angular momentum content, therefore can attain surface rotational velocities close to critical throughout the whole MS phase. 

In the following subsection, we extend the work by \citet{Haemmerle2013} to lower metallicities by studying the impact of different initial rotation profiles on the evolution
of a  9 M$_{\sun}$ at Z=0.002. 

\subsection{Computing the stellar structure at the ZAMS for Z=0.002}

We first computed a model with the accretion phase during pre-MS evolution.
As in \citet{Haemmerle2013}, we start with a fully convective 0.7 M$_{\sun}$ hydrostatic core
at the top of the Hayashi line ($Z=0.002$, $T_{\rm eff}=4427[K]$ and $L=10.3L_{\sun}$).
We assumed that the star accretes mass at a rate $\dot M$ given by
\begin{equation}
\log2\dot M=-5.28+\log\frac{L}{L_{\sun}}\cdot(0.752-0.0278\cdot\log\frac{L}{L_{\sun}}),
\end{equation}
where $L$ is the bolometric luminosity of the star.
This mass accretion rate, first used by \citet{Behrend2001}, was inspired by the observational correlation 
between mass outflows and bolometric luminosities in ultra-compact HII regions
\citep{Churchwell1999,Henning2000} and from a fit with observed Herbig Ae/Be stars \citep{Behrend2001}.
Because the luminosity of the star changes during accretion, the mass accretion rate is time dependent, and
except for very early in the pre-MS evolution, it increases with the age.
This rate does not depend on the final mass of the star.  
In this scenario, all stars instead accrete at the same rate until they reach their final mass $M_{\rm ZAMS}$ and
once accretion is supressed, the star evolves at constant mass.

At each timestep during the accretion phase we assigned the same values for thermodynamic quantities
to the material that is accreted as to the material of the surface of the stellar model in the previous timestep.
This choice corresponds to the \textit{cold-disc accretion} scenario.
The same was done for the angular velocity: at each time step, we assigned 
the same value of $\Omega$ to the material that is accreted as to the surface layers of the model in the previous timestep.
This choice determines the angular momentum accretion law $\dot J$, 
and this law corresponds to a specific angular momentum in the accreted material that is almost constant with time.

With such a law, the total angular momentum of the star on the ZAMS for a given mass and metallicity is completely determined 
by the angular velocity we give to the initial model.
We arbitrarily took $\Omega_{\rm ini}=2\times10^{-5}$ s$^{-1}$ and considered M$_{\rm ZAMS}$=9 M$_{\sun}$.
This leads to a total angular momentum contained in the star of $J_{\rm tot}=9.59\times10^{51}\,\rm[g\,cm^2\,s^{-1}]$
when the accretion phase ends.

We computed two additional models with M=9 M$_{\sun}$ and $Z=0.002$ starting the ZAMS with this same value of $J_{\rm tot}$,
one with a flat internal rotation profile $\nabla\Omega=0$ and
one with an ad hoc rotation profile, steeper than the one we obtain after pre-MS evolution: $\nabla\Omega>\nabla\Omega_{\rm PMS}$.

\subsection{Impact of the initial rotational profile for Z=0.002} \label{initialZ}

\begin{figure}
\includegraphics[angle=270,width=0.49\textwidth]{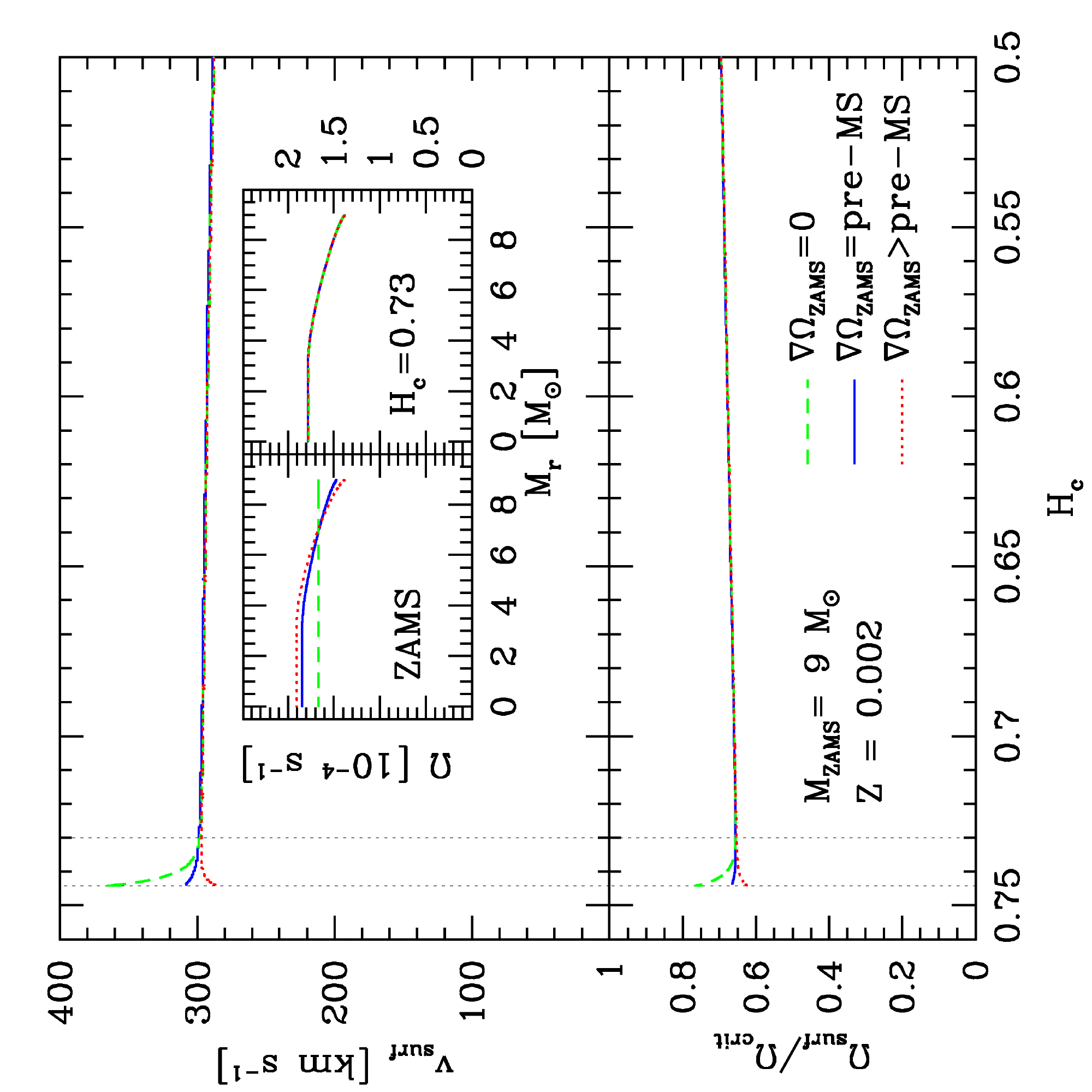}
\caption{Evolution of the stellar rotation during the first part of the MS
for the three models with 9{\rm M$_{\sun}$} and $Z=0.002$
with the same value of $J_{\rm tot}$, but with different rotation profiles on the ZAMS,
as described in the text.
\textit{Upper panel:} Equatorial rotational velocity 
as a function of the central mass fraction of hydrogen $H_c$.
The internal rotation profiles on the ZAMS ($H_c\simeq0.744$) and at $H_c=0.73$ are superimposed.
\textit{Lower panel:} Ratio of the surface angular velocity to the critical angular velocity
as a function of the central mass fraction of hydrogen.
In both panels, the two stages of the rotation profiles are indicated by vertical dotted lines.}
\label{be8_v}
\end{figure}

The evolution of the rotational velocity profiles for the three models described in the previous subsection
during the first part of the MS is plotted in Fig. \ref{be8_v}. In the upper panel of this figure we also show the three internal rotation profiles
on the ZAMS (for a central mass fraction of hydrogen $H_c\simeq0.744$)
and at $H_c=0.73$, which corresponds to an age of $\sim4\%$ of the total MS duration.
This figure shows that the model with the flat internal rotation profile ($\nabla\Omega_{\rm ZAMS}=0$) 
begins its MS evolution with a sharp decrease of $v_{\rm surf}$.
The strong meridional currents quickly modify the artificial flat rotation profile, producing a
decrease of the surface velocity and an increase of the angular velocity in the internal regions.
The same occurs with the model obtained through pre-MS calculations:
despite having a differential rotation on the ZAMS, the rotation profile is not in a stationary situation either, 
and meridional currents act to increase $\nabla\Omega$ and decrease the surface velocity as for $\nabla\Omega_{\rm ZAMS}=0$,
but in a weaker fashion, since the rotation profile is closer to a quasi-equilibrium configuration.
The third model we propose at the ZAMS has an internal profile that is steeper than the as for the pre-MS calculations we described before.
For this ad hoc profile, the convergence to the quasi-equilibrium configuration (hereafter {\it initial MS rotation profile}) 
corresponds to an increase of the surface velocity and a decrease of the angular velocity in the central regions of the star.
Fig. \ref{be8_v} shows that very shortly after the ZAMS (4\% of the MS duration), 
the rotation profile of the three models with the same angular momentum content has
already converged to the same initial MS rotation profile.
After this stage, all models evolve identically in terms of surface velocity and evolutionary tracks on the HRD.

\begin{figure}
\includegraphics[angle=270,width=0.49\textwidth]{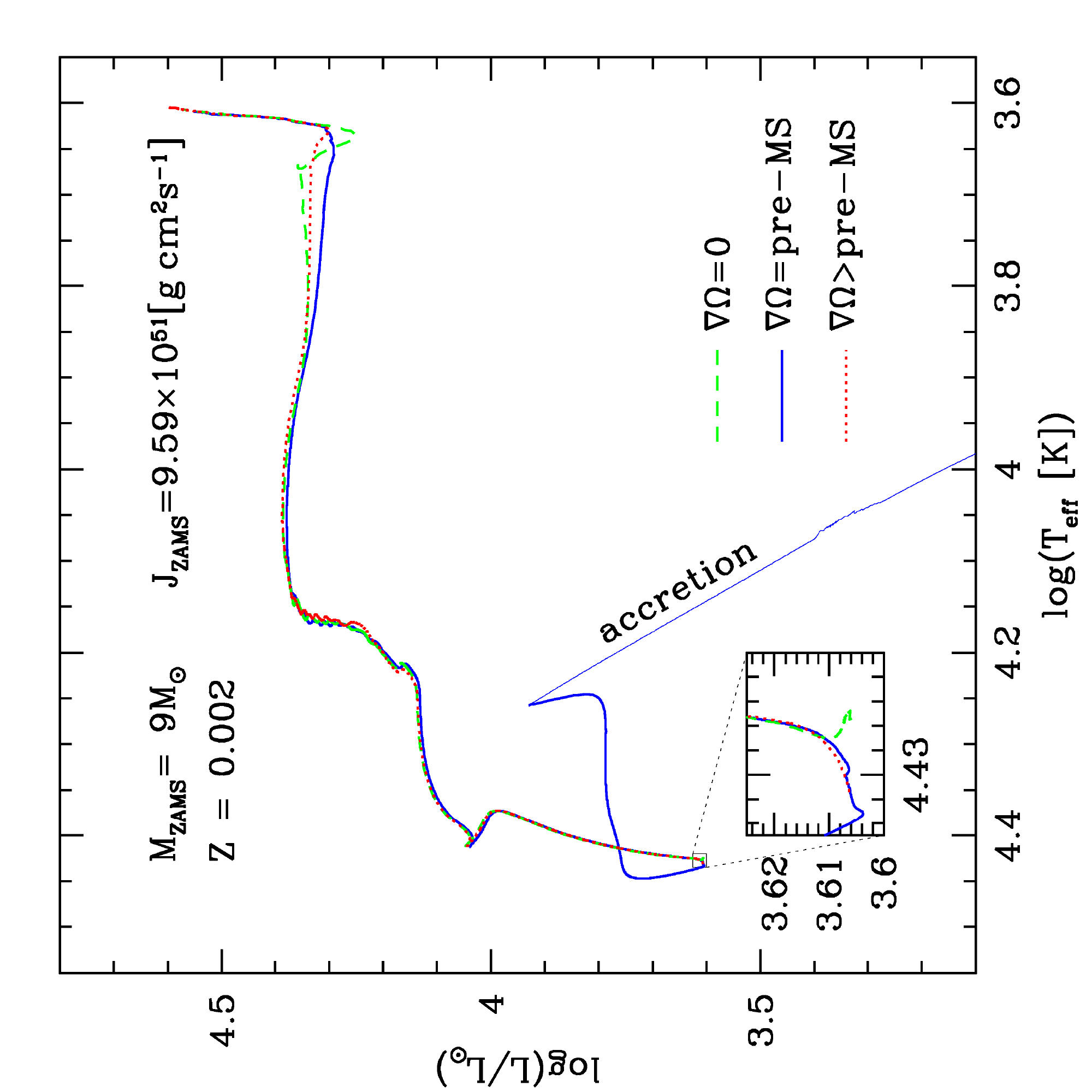}
\caption{HR diagram of the three models with 9{\rm M$_{\sun}$} and $Z=0.002$
with the same value of $J_{\rm tot}$, but with different rotation profiles on the ZAMS,
as described in the text. For the model computed with the pre-MS,
the part of the track corresponding to the accretion phase is indicated.}
\label{be8_hr}
\end{figure}

The evolutionary tracks on the HRD of the models with the three initial conditions are shown in Fig. \ref{be8_hr} together
with a zoom on the ZAMS showing that the three different tracks converge to the same.
There are only small differences in the evolutionary tracks after the MS phase when crossing to the red part of the HRD after the end of central He burning.
This is because the values of $J_{\rm tot}$ in the three models are not exactly the same.
The small differences in $J_{\rm tot}$ ($\simeq0.05\%$) can produce these differences on the HR tracks
in these stages that are very sensitive to the rotation properties of the star.

The behaviour of the equatorial rotational velocity just after the ZAMS for the different initial internal structures 
illustrates the reason why we need to propose an ad hoc rotation profile on the ZAMS in some cases to obtain models
with a surface velocity close to the critical velocity during the whole main sequence.
As shown by \citet{Haemmerle2013} for $Z=0.014$, models with a pre-MS evolution that have M$_{\rm ZAMS}\geq$6 M$_{\sun}$
begin their MS evolution with a decrease of $\rm v_{surf}/v_{crit}$,
as for $\nabla\Omega_{\rm ZAMS}=0$, but in a weaker way.
For $Z=0.002$, this initial decrease of $\rm v_{surf}/v_{crit}$ occurs for the whole mass range we considered here.
Where the decrease in the equatorial velocity is strong enough, a limitation in $J_{\rm tot}$ is imposed, of the same 
type as for $\nabla\Omega_{\rm ZAMS}=0$, so that an ad hoc steeper internal rotational profile is required at the ZAMS to be able to generate 
stellar models with higher values of $J_{\rm tot}$. These high initial values of angular momentum content are necessary for a rotation of the star 
with $\rm v_{surf}/v_{crit}$ close to one during the whole MS.

We have shown that the MS evolution of models with a steep ad hoc internal rotation profile on the ZAMS
is virtually the same as the MS evolution of models with $\nabla\Omega_{\rm ZAMS}=0$.
We used these ad hoc rotation profiles on the ZAMS when the rotation profile obtained from pre-MS evolution
did not allow us to obtain a $\rm v_{surf}/v_{crit}$ high enough at the beginning of the MS.

The question that naturally arises is whether there is a pre-MS scenario ($\dot M$, $\dot J$), different from the one proposed here, that produces 
such a steep rotation profile on the ZAMS. Answering this question is beyond the scope of this article, therefore we leave this 
to be discussed in a forthcoming work.

Our findings are also useful because they show that models with a solid rotation profile at the ZAMS can represent the evolution of a star with the same initial angular momentum content, but any initial internal rotation profile, except at the very beginning of the MS. We should bear in mind, however, that
 the surface equatorial velocity when the \textit{initial MS rotation profile} is reached corresponds to a lower value than the equatorial velocity 
for the solidly rotating model at the ZAMS. 
 
In this sense, each model presented by \citet{Georgy2013a} would adequately represent a differentially rotating
star at the ZAMS with an initial rotational velocity rate $\omega$ corresponding to the rate obtained when the \textit{initial MS rotation profile}
is reached, which corresponds to the lowest rotational rate achieved soon after the ZAMS, when the star 
reaches a quasi-equilibrium state, as seen in Fig. \ref{be8_v}.

Because we intend to understand the evolution of very rapidly rotating stars, we saught to obtain stellar evolution models with rotational rates close to critical 
during the whole MS. We present here rapidly rotating stellar models with a large angular momentum content and 
 a non-flat rotational profile at the ZAMS that can soon reach the critical limit. 

In the following section we compare these new stellar evolution models with 
large angular momentum content with the results reported by \citet{Georgy2013a}.

\section{Results}
Following what has been presented in Section 2 we can assume that each model computed asuming solid-body rotation and 
different rotational rate at the ZAMS \citep{Georgy2013a} 
evolves identically to models with the same mass, metallicity, and angular momentum content, regardless of the internal rotational profile, 
except for very early in the main sequence. 
 
Because we intend to have a grid of models with initial rotation rates ranging from zero to almost critically rotating in order 
to be able to study the evolution of rotating stellar populations, we still need to complete the range of models for stars that
rotate close to the critical limit since early in the MS. To do this, we generated models with a non-solid-body rotation profile at the ZAMS and
 a larger angular momentum content than those presented in \citet{Georgy2013a}. As explained in Section 2, for models that failed to reach these large
angular momentum content through the proposed accretion scenario, an ad hoc initial rotational profile was proposed.
 
In this way, we produced new evolutionary tracks with initial rotational rates $\omega$ at the ZAMS higher than 0.9. 
In this section, we describe our new results and compare them with the grids by \citet{Georgy2013a}. 
The electronic tables for these evolutionary tracks are made available at http://obswww.unige.ch/Recherche/evol/-Database-.

{ Tables \ref{TabListModelsZ014}, \ref{TabListModelsZ006}, and \ref{TabListModelsZ002} show the overall characteristics at the ZAMS and at the end
of the main burning stages for models with different masses and angular momentum content at Z=0.014, Z=0.006, and Z=0.002. The format of the tables is similar
to the format presented by \citet{Georgy2013a}, except that we provide here the initial angular momentum instead of the mean rotational velocity through the MS 
 to make the comparison among the models straightforward. 
Because at the end of the He and C burning the surface velocities are low, the rotation period and the ratio between the 
surface and the core angular velocities are listed in these tables. As for red giant stars \citep{Eggenberger2012}, it is also possible to estimate this ratio from asteroseismic observations for more massive stars, which 
can set constraints on the different angular momentum transfer prescriptions in stellar evolution models.
 The rows that list to our new calculations with a large angular momentum content are indicated with an asterisk. 
The remaining rows list the older models by \citet{Georgy2013a}, but their initial rotational velocity and rotational rate have been corrected assuming that 
these quantities correspond to the \textit{initial MS rotation profile}, as described in Sect. 2. The modified quantities are indicated with the symbol $\dagger$.}

\begin{figure}
\centerline{\includegraphics[trim = 0mm 10mm 0mm 20mm, clip, width=0.5\textwidth, angle=0]{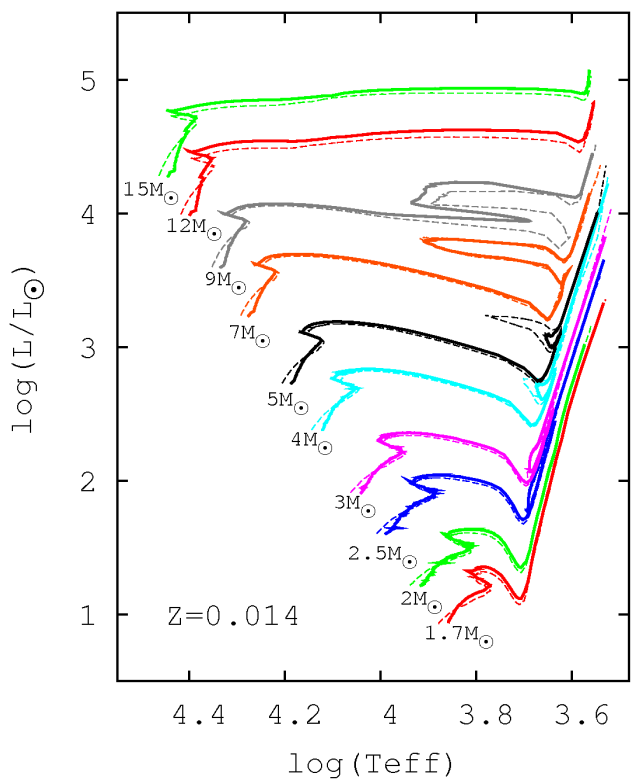}}
\caption{HRD of our models with 1.7 (red), 2 (green), 2.5 (blue), 3 (magenta), 4 (cyan), 5 (black), 7 (orange), 9 (grey), 12 (red) and 15 (green) solar masses corresponding to the largest angular momentum content 
(thick continuous lines) and those with the highest rotational rate at the ZAMS presented by \citet{Georgy2013a} (thin dashed lines), for solar metallicity.}
\label{HRD_14}
\end{figure}

\begin{figure*}
\centerline{\includegraphics[trim = 0mm 10mm 0mm 20mm, clip,width=0.5\textwidth, angle=0]{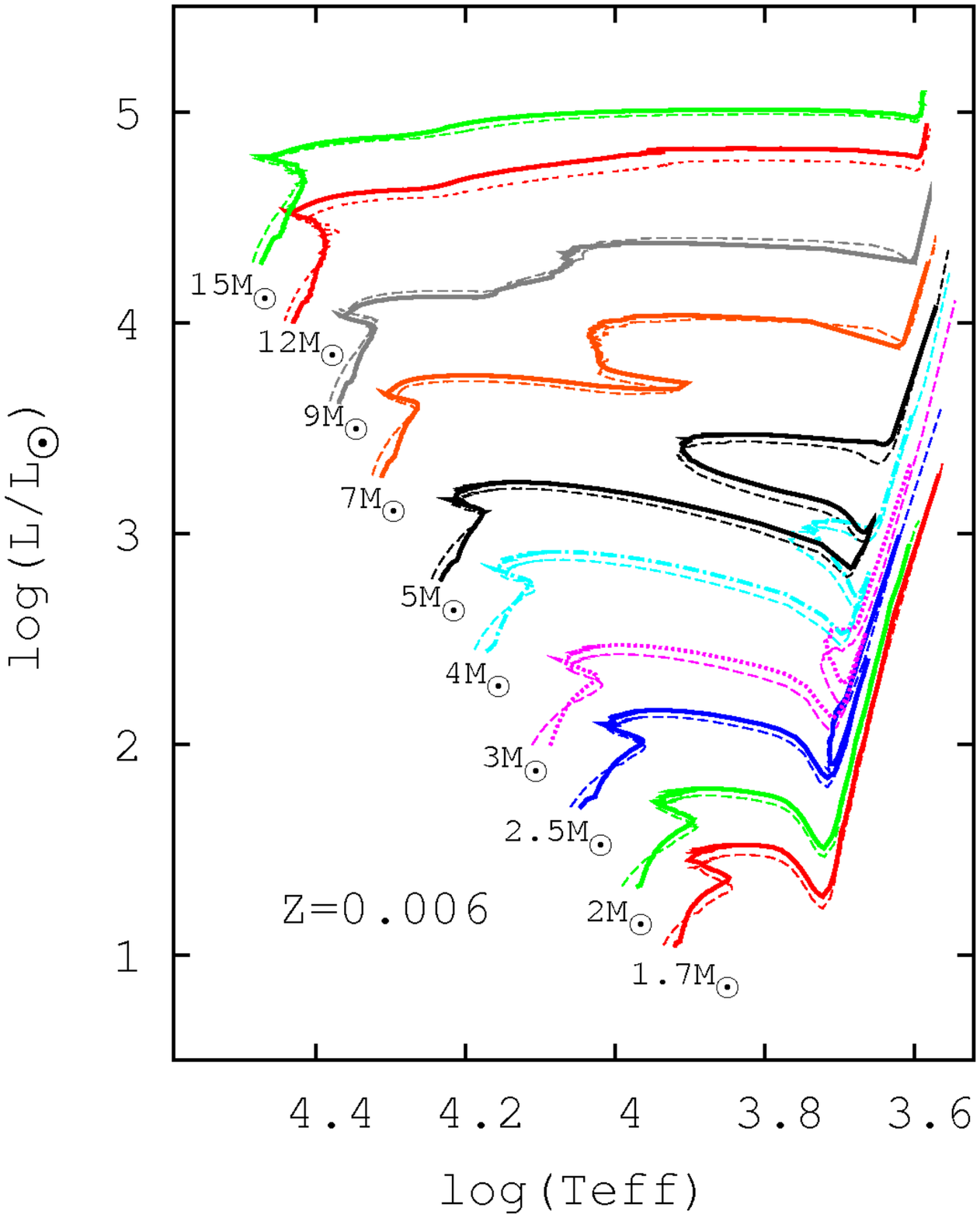}
\includegraphics[trim = 0mm 10mm 0mm 20mm, clip,width=0.5\textwidth, angle=0]{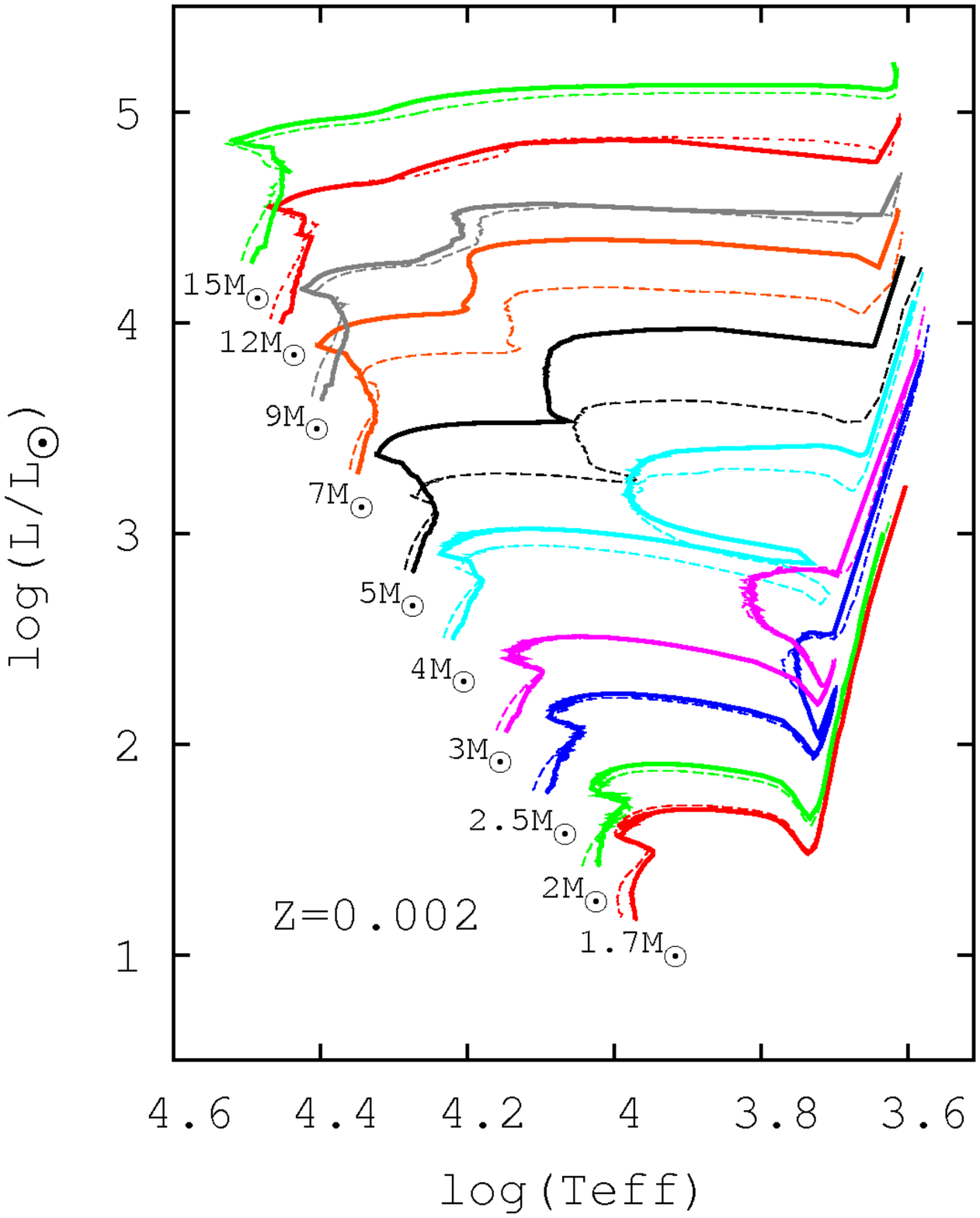}}
\caption{Same as Fig. \ref{HRD_14}, but for Z=0.006 (left) and Z=0.002 (right).}
\label{HRD_02}
\end{figure*}

Figures \ref{HRD_14} and \ref{HRD_02}  show the location in the HRD of the models with the largest angular momentum content 
we computed for each mass and metallicity, together with the models with the fastest rotation at the ZAMS presented previously by \citet{Georgy2013a}. 
In the following subsections, we discuss the most remarkable 
differences and similitudes between the previously computed and the new models.

\begin{sidewaystable*}
\caption{Main parameters of A-B stars at $Z=0.014=Z_{\sun}$}
\centering
\scalebox{0.63}{\begin{tabular}{rccc||ccccccc|ccccccc|ccccccc}
\hline\hline
\multicolumn{4}{c||}{} & \multicolumn{7}{c|}{End of H-burning} & \multicolumn{7}{c|}{End of He-burning} & \multicolumn{7}{c}{End of C-burning}\\
     $M_\text{ini}$ & $\Omega/\Omega_\text{crit, ini}$ & $v_\text{eq}$ & $\mathcal{L}$ & $\tau_\text{H}$ & $M$ & $v_\text{eq}$ & $v_\text{eq}/v_\text{crit}$ & $Y_\text{surf}$ & $\text{N}/\text{C}$ & $\text{N}/\text{O}$ & $\tau_\text{He}$ & $M$ & $P_\text{rot}$ & $\Omega_\text{cen}/\Omega_\text{sur}$ & $Y_\text{surf}$ & $\text{N}/\text{C}$ & $\text{N}/\text{O}$ & $\tau_\text{C}$ & $M$ & $P_\text{rot}$ & $\Omega_\text{cen}/\Omega_\text{sur}$ & $Y_\text{surf}$ & $\text{N}/\text{C}$ & $\text{N}/\text{O}$ \\
$M_{\sun}$ & &km s$^{-1}$ &  $10^{51}$\,g\,cm$^2$\,s$^{-1}$  & Myr & $M_{\sun}$ & km s$^{-1}$ & & \multicolumn{3}{c|}{mass fract.} & Myr & $M_{\sun}$ & day & & \multicolumn{3}{c|}{mass fract.} & kyr & $M_{\sun}$ & day & & \multicolumn{3}{c}{mass fract.}\\
\hline
     $ 15.00$ & $0.00$ & $  0.$ & $^{\dagger}0.$ & $   11.015$ & $ 14.81$ & $  0.$ & $0.000$ & $0.2660$ & $  0.2885$ & $  0.1152$ & $    1.315$ & $ 13.34$ & --                  & --                  & $0.3054$ & $  2.0811$ & $  0.5211$ & $    4.721$ & $ 13.25$ & --                  & --                  & $0.3390$ & $  3.0660$ & $  0.7360$ \\
     $ 15.00$ & $^{\dagger}0.43$ & $^{\dagger}188.$ & $^{\dagger}16.11$ & $   13.399$ & $ 14.71$ & $139.$ & $0.329$ & $0.2835$ & $  1.7907$ & $  0.4568$ & $    1.365$ & $ 11.19$ & $7.856\cdot 10^{5}$ & $2.153\cdot 10^{7}$ & $0.3719$ & $  5.5490$ & $  0.9928$ & $    2.121$ & $ 11.07$ & $1.106\cdot 10^{6}$ & $2.685\cdot 10^{9}$ & $0.3954$ & $  6.5423$ & $  1.1182$ \\
     $ 15.00$ & $^{\dagger}0.81$ & $^{\dagger}400.$ & $^{\dagger}30.90$ & $   14.189$ & $ 14.66$ & $287.$ & $0.639$ & $0.3199$ & $  8.1555$ & $  0.8963$ & $    1.441$ & $ 12.63$ & $5.908\cdot 10^{4}$ & $2.815\cdot 10^{6}$ & $0.4024$ & $ 17.4288$ & $  1.3403$ & $    2.543$ & $ 12.51$ & $9.546\cdot 10^{4}$ & $4.241\cdot 10^{8}$ & $0.4208$ & $ 20.0816$ & $  1.4607$ \\
    *$ 15.00$& $0.95$& $512.$ & $36.13$ & $   15.675$ & $ 14.56$ & $278.$ & $0.629$ & $0.3424$ & $ 12.4722$ & $  1.0927$ & $    1.302$ & $ 11.96$ & $8.678\cdot 10^{4}$ & $3.939\cdot 10^{6}$ & $0.4338$ & $ 25.3205$ & $  1.6061$ & $    2.302$ & $ 11.83$ & $1.487\cdot 10^{5}$ & $6.112\cdot 10^{8}$ & $0.4502$ & $ 28.6042$ & $  1.7222$ \\
\hline
     $ 12.00$ & $0.00$ & $  0.$ & $  0.$ & $   15.331$ & $ 11.94$ & $  0.$ & $0.000$ & $0.2660$ & $  0.2885$ & $  0.1152$ & $    2.124$ & $ 11.36$ & --                  & --                  & $0.3007$ & $  1.8374$ & $  0.4918$ & $    7.858$ & $ 11.31$ & --                  & --                  & $0.3072$ & $  1.9788$ & $  0.5280$ \\
     $ 12.00$ & $^{\dagger}0.44$ & $^{\dagger}178.$ & $^{\dagger}10.54$ & $   18.540$ & $ 11.91$ & $164.$ & $0.397$ & $0.2755$ & $  1.4363$ & $  0.3818$ & $    1.892$ & $ 10.40$ & $8.402\cdot 10^{4}$ & $2.809\cdot 10^{6}$ & $0.3272$ & $  4.4674$ & $  0.7804$ & $    6.134$ & $ 10.31$ & $1.327\cdot 10^{5}$ & $5.280\cdot 10^{8}$ & $0.3502$ & $  5.3748$ & $  0.8895$ \\
     $ 12.00$ & $^{\dagger}0.82$ & $^{\dagger}385.$ & $^{\dagger}20.09$ & $   19.302$ & $ 11.89$ & $333.$ & $0.748$ & $0.3071$ & $  9.0862$ & $  0.8352$ & $    2.357$ & $ 10.80$ & $3.044\cdot 10^{4}$ & $1.837\cdot 10^{6}$ & $0.3552$ & $ 18.2425$ & $  1.0840$ & $    0.066$ & $ 10.73$ & $4.571\cdot 10^{4}$ & $5.457\cdot 10^{7}$ & $0.3780$ & $ 21.0351$ & $  1.1874$ \\
     *$ 12.00$& $0.95$      & $494.$ & $23.33$  & $   21.675$ & $ 11.83$ & $323.$ & $0.739$ & $0.3284$ & $ 14.1578$ & $  1.0083$ & $    2.028$ & $ 10.53$ & $4.182\cdot 10^{4}$ & $2.267\cdot 10^{6}$ & $0.3772$ & $ 25.4539$ & $  1.2692$ & $    0.069$ & $ 10.45$ & $5.753\cdot 10^{4}$ & $5.759\cdot 10^{7}$ & $0.4010$ & $ 29.2844$ & $  1.3905$\\
\hline     
     $  9.00$ & $0.00$ & $  0.$ & $  0.$ & $   26.262$ & $  8.99$ & $  0.$ & $0.000$ & $0.2660$ & $  0.2885$ & $  0.1152$ & $    3.492$ & $  8.80$ & --                  & --                  & $0.2824$ & $  1.6090$ & $  0.4219$ & & & & & & & \\
     $  9.00$ & $^{\dagger}0.44$ & $^{\dagger}170.$ & $^{\dagger}5.98$ & $   31.464$ & $  8.99$ & $170.$ & $0.427$ & $0.2709$ & $  1.0243$ & $  0.3023$ & $    3.351$ & $  8.76$ & $1.001\cdot 10^{4}$ & $3.890\cdot 10^{5}$ & $0.3285$ & $  4.7700$ & $  0.7896$ & $    0.009$ & $  8.72$ & $3.673\cdot 10^{4}$ & $7.449\cdot 10^{7}$ & $0.3290$ & $  4.9811$ & $  0.8029$ \\
     $  9.00$ & $^{\dagger}0.83$ & $^{\dagger}364.$ & $^{\dagger}11.40$ & $   32.482$ & $  8.98$ & $380.$ & $0.888$ & $0.2962$ & $  7.0486$ & $  0.7520$ & $    4.610$ & $  8.73$ & $9.626\cdot 10^{3}$ & $7.050\cdot 10^{5}$ & $0.3296$ & $ 15.9175$ & $  0.9689$ & & & & & & & \\
     *$  9.00$& $0.95$      & $464.$ & $13.09$ & $   33.120$ & $  8.94$ & $391.$ & $0.903$ & $0.3098$ & $ 11.9060$ & $  0.8770$ & $    4.250$ & $  8.80$ & $8.105\cdot 10^{3}$ & $5.597\cdot 10^{5}$ & $0.3465$ & $ 25.2410$ & $  1.0864$ & & & & & & & \\
\hline
     $  7.00$ & $0.00$ & $  0.$ & $  0.$ & $   41.722$ & $  7.00$ & $  0.$ & $0.000$ & $0.2660$ & $  0.2885$ & $  0.1152$ & $    6.919$ & $  6.92$ & --                  & --                  & $0.2792$ & $  1.5177$ & $  0.3964$ & & & & & & & \\
     $  7.00$ & $^{\dagger}0.44$ & $^{\dagger}161.$ & $^{\dagger}3.59$ & $   51.151$ & $  7.00$ & $160.$ & $0.420$ & $0.2685$ & $  0.7166$ & $  0.2341$ & $    6.795$ & $  6.89$ & $4.772\cdot 10^{3}$ & $2.193\cdot 10^{5}$ & $0.3190$ & $  4.0102$ & $  0.7192$ & & & & & & & \\
     $  7.00$ & $^{\dagger}0.83$ & $^{\dagger}347.$ & $^{\dagger}6.87$ & $   52.900$ & $  6.99$ & $362.$ & $0.892$ & $0.2871$ & $  4.3773$ & $  0.6282$ & $    8.465$ & $  6.89$ & $2.508\cdot 10^{3}$ & $2.187\cdot 10^{5}$ & $0.3291$ & $ 13.0544$ & $  0.9246$ & & & & & & & \\
     *$  7.00$& $0.95$       & $442.$ & $7.79$   & $   54.258$ & $  6.96$ & $360.$ & $0.882$ & $0.2984$ & $  7.3498$ & $  0.7410$ & $    8.674$ & $  6.87$ & $3.501\cdot 10^{3}$ & $2.884\cdot 10^{5}$ & $0.3388$ & $ 18.3308$ & $  0.9904$  \\
\cline{1-18}
     $  5.00$ & $0.00$ & $  0.$ & $  0.$ & $   88.195$ & $  5.00$ & $  0.$ & $0.000$ & $0.2660$ & $  0.2885$ & $  0.1152$ & $   19.380$ & $  4.96$ & --                  & --                  & $0.2792$ & $  1.4819$ & $  0.3864$ & & & & & & & \\
     $  5.00$ & $^{\dagger}0.44$ & $^{\dagger}150.$ & $^{\dagger}1.80$ & $  109.085$ & $  5.00$ & $143.$ & $0.409$ & $0.2670$ & $  0.4995$ & $  0.1781$ & $   16.800$ & $  4.95$ & $1.416\cdot 10^{3}$ & $7.935\cdot 10^{4}$ & $0.3154$ & $  3.3394$ & $  0.6627$ & & & & & & & \\
     $  5.00$ & $^{\dagger}0.83$ & $^{\dagger}325.$ & $^{\dagger}3.44$ & $  113.858$ & $  4.99$ & $316.$ & $0.866$ & $0.2771$ & $  2.1374$ & $  0.4527$ & $   21.160$ & $  4.94$ & $8.364\cdot 10^{2}$ & $8.189\cdot 10^{4}$ & $0.3256$ & $  8.1090$ & $  0.8382$ & & & & & & & \\
     *$  5.00$& $0.95$      & $419.$ & $3.90$ & $  117.370$ & $  4.98$ & $316.$ & $0.859$ & $0.2891$ & $  4.3180$ & $  0.6092$ & $   19.920$ & $  4.93$ & $1.063\cdot 10^{3}$ & $1.023\cdot 10^{5}$ & $0.3327$ & $ 10.9402$ & $  0.8951$  \\
\cline{1-18}
     $  4.00$ & $0.00$ & $  0.$ & $  0.$ & $  152.088$ & $  4.00$ & $  0.$ & $0.000$ & $0.2660$ & $  0.2885$ & $  0.1152$ & $   37.750$ & $  3.97$ & --                  & --                  & $0.2823$ & $  1.5023$ & $  0.3912$ & & & & & & & \\
     $  4.00$ & $^{\dagger}0.45$ & $^{\dagger}142.$ & $^{\dagger}1.13$ & $  187.477$ & $  4.00$ & $129.$ & $0.399$ & $0.2666$ & $  0.4202$ & $  0.1554$ & $   34.852$ & $  3.97$ & $8.134\cdot 10^{2}$ & $5.097\cdot 10^{4}$ & $0.3167$ & $  3.0686$ & $  0.6380$ & & & & & & & \\
     $  4.00$ & $^{\dagger}0.84$ & $^{\dagger}310.$ & $^{\dagger}2.16$ & $  198.351$ & $  3.99$ & $282.$ & $0.849$ & $0.2743$ & $  1.5901$ & $  0.3846$ & $   37.370$ & $  3.96$ & $3.954\cdot 10^{2}$ & $4.232\cdot 10^{4}$ & $0.3276$ & $  6.4637$ & $  0.7997$ & & & & & & & \\
     *$  4.00$& $0.95$      & $408.$ & $2.47$  & $  205.350$ & $  3.97$ & $297.$ & $0.897$ & $0.2859$ & $  3.2599$ & $  0.5439$ & $   36.264$ & $  3.94$ & $5.206\cdot 10^{2}$ & $5.311\cdot 10^{4}$ & $0.3359$ & $  8.8096$ & $  0.8687$  \\
\cline{1-18}
     $  3.00$ & $0.00$ & $  0.$ & $  0.$ & $  320.600$ & $  3.00$ & $  0.$ & $0.000$ & $0.2660$ & $  0.2885$ & $  0.1152$ & $  117.109$ & $  2.99$ & --                  & --                  & $0.2868$ & $  1.5318$ & $  0.3928$ & & & & & & & \\
     $  3.00$ & $^{\dagger}0.45$ & $^{\dagger}133.$ & $^{\dagger}0.62$ & $  400.944$ & $  3.00$ & $116.$ & $0.395$ & $0.2664$ & $  0.3698$ & $  0.1402$ & $   97.083$ & $  2.98$ & $2.641\cdot 10^{2}$ & $2.010\cdot 10^{4}$ & $0.3213$ & $  2.8644$ & $  0.6132$ & & & & & & & \\
     $  3.00$ & $^{\dagger}0.84$ & $^{\dagger}290.$ & $^{\dagger}1.19$ & $  422.412$ & $  3.00$ & $225.$ & $0.748$ & $0.2719$ & $  1.1641$ & $  0.3145$ & $  105.941$ & $  2.98$ & $1.928\cdot 10^{2}$ & $2.302\cdot 10^{4}$ & $0.3314$ & $  5.3924$ & $  0.7517$ & & & & & & & \\
     *$  3.00$& $0.95$      & $372.$ & $1.36$ & $  441.770$ & $  2.98$ & $239.$ & $0.797$ & $0.2805$ & $  2.2412$ & $  0.4474$ & $  97.862$ & $  2.96$ & $2.598\cdot 10^{2}$ & $2.718\cdot 10^{4}$ & $0.3406$ & $  7.0227$ & $  0.8219$ \\
\cline{1-18}
     $  2.50$ & $0.00$ & $  0.$ & $  0.$ & $  537.964$ & $  2.50$ & $  0.$ & $0.000$ & $0.2660$ & $  0.2885$ & $  0.1152$ & $  236.764$ & $  2.49$ & --                  & --                  & $0.2852$ & $  1.4444$ & $  0.3662$ & & & & & & & \\
     $  2.50$ & $^{\dagger}0.45$ & $^{\dagger}128.$ & $^{\dagger}0.43$ & $  665.370$ & $  2.50$ & $109.$ & $0.389$ & $0.2663$ & $  0.3491$ & $  0.1338$ & $  185.800$ & $  2.49$ & $2.206\cdot 10^{2}$ & $1.648\cdot 10^{4}$ & $0.3206$ & $  2.6893$ & $  0.5737$ & & & & & & & \\
     $  2.50$ & $^{\dagger}0.83$ & $^{\dagger}277.$ & $^{\dagger}0.82$ & $  704.202$ & $  2.50$ & $228.$ & $0.800$ & $0.2711$ & $  1.0004$ & $  0.2816$ & $  194.550$ & $  2.48$ & $1.478\cdot 10^{2}$ & $1.814\cdot 10^{4}$ & $0.3301$ & $  4.7606$ & $  0.6958$ & & & & & & & \\
     *$  2.50$& $0.95$     & $348.$ & $0.94$ & $  745.024$ & $  2.49$ & $200.$ & $0.755$ & $0.2755$ & $  1.5805$ & $  0.3629$ & $  195.844$ & $  2.47$ & $1.923\cdot 10^{2}$ & $1.867\cdot 10^{4}$ & $0.3413$ & $  6.1237$ & $  0.7648$ \\
\cline{1-18}
     $  2.00$ & $0.00$ & $  0.$ & $  0.$ & $ 1008.888$ & $  2.00$ & $  0.$ & $0.000$ & $0.2660$ & $  0.2885$ & $  0.1152$ \\
     $  2.00$ & $^{\dagger}0.46$ & $^{\dagger}125.$ & $^{\dagger}0.28$ & $ 1267.754$ & $  2.00$ & $105.$ & $0.392$ & $0.2664$ & $  0.3352$ & $  0.1294$ \\
     $  2.00$ & $^{\dagger}0.83$ & $^{\dagger}258.$ & $^{\dagger}0.51$ & $ 1340.230$ & $  2.00$ & $210.$ & $0.775$ & $0.2704$ & $  0.7895$ & $  0.2373$ \\
     *$  2.00$& $0.96$     & $337.$ & $0.60$       & $ 1427.870$ & $  1.99$ & $189.$ & $0.750$ & $0.2756$ & $  1.3715$ & $  0.3270$ \\
\cline{1-11}
     $  1.70$ & $0.00$ & $  0.$ & $0$ & $ 1633.971$ & $  1.70$ & $  0.$ & $0.000$ & $0.2660$ & $  0.2885$ & $  0.1152$ \\
     $  1.70$ & $^{\dagger}0.46$ & $^{\dagger}120.$ & $^{\dagger}0.20$ & $ 2068.535$ & $  1.70$ & $102.$ & $0.391$ & $0.2664$ & $  0.3214$ & $  0.1252$ \\
     $  1.70$ & $^{\dagger}0.85$ & $^{\dagger}255.$ & $^{\dagger}0.39$ & $ 2181.676$ & $  1.70$ & $212.$ & $0.809$ & $0.2711$ & $  0.7294$ & $  0.2228$ \\
     *$  1.70$& $0.95$      & $311.$ & $0.44$ & $ 2386.735$ & $  1.69$ & $214.$ & $0.833$ & $0.2805$ & $  1.5606$ & $  0.3403$ \\
\hline
\hline
\end{tabular}}
\label{TabListModelsZ014}
\end{sidewaystable*}

\begin{sidewaystable*}
\caption{Main parameters of A-B stars at $Z=0.006=Z_\text{LMC}$}
\centering
\scalebox{0.63}{\begin{tabular}{rccc||ccccccc|ccccccc|ccccccc}
\hline\hline
\multicolumn{4}{c||}{} & \multicolumn{7}{c|}{End of H-burning} & \multicolumn{7}{c|}{End of He-burning} & \multicolumn{7}{c}{End of C-burning}\\
     $M_\text{ini}$ & $\Omega/\Omega_\text{crit, ini}$ & $v_\text{eq}$ & $\mathcal{L}$ & $\tau_\text{H}$ & $M$ & $v_\text{eq}$ & $v_\text{eq}/v_\text{crit}$ & $Y_\text{surf}$ & $\text{N}/\text{C}$ & $\text{N}/\text{O}$ & $\tau_\text{He}$ & $M$ & $P_\text{rot}$ & $\Omega_\text{cen}/\Omega_\text{sur}$ & $Y_\text{surf}$ & $\text{N}/\text{C}$ & $\text{N}/\text{O}$ & $\tau_\text{C}$ & $M$ & $P_\text{rot}$ & $\Omega_\text{cen}/\Omega_\text{sur}$ & $Y_\text{surf}$ & $\text{N}/\text{C}$ & $\text{N}/\text{O}$ \\
$M_{\sun}$ & &km s$^{-1}$ &  $10^{51}$\,g\,cm$^2$\,s$^{-1}$& Myr & $M_{\sun}$ & km s$^{-1}$ & & \multicolumn{3}{c|}{mass fract.} & Myr & $M_{\sun}$ & day & & \multicolumn{3}{c|}{mass fract.} & kyr & $M_{\sun}$ & day & & \multicolumn{3}{c}{mass fract.}\\
\hline
     $ 15.00$ & $0.00$ & $  0.$ & $  0.$ & $   11.097$ & $ 14.87$ & $  0.$ & $0.000$ & $0.2559$ & $  0.2885$ & $  0.1152$ & $    1.338$ & $ 14.00$ & --                  & --                  & $0.2685$ & $  1.3634$ & $  0.3566$ & $    3.504$ & $ 13.90$ & --                  & --                  & $0.3214$ & $  2.7453$ & $  0.6739$ \\
     $ 15.00$ & $^{\dagger}0.43$ & $^{\dagger}195.$ & $^{\dagger}16.60$ & $   13.409$ & $ 14.80$ & $168.$ & $0.370$ & $0.2776$ & $  2.7679$ & $  0.5921$ & $    1.286$ & $ 13.92$ & $3.766\cdot 10^{4}$ & $1.143\cdot 10^{6}$ & $0.3490$ & $  6.5843$ & $  1.0305$ & $    2.475$ & $ 13.80$ & $6.487\cdot 10^{4}$ & $1.772\cdot 10^{8}$ & $0.3845$ & $  8.8557$ & $  1.2540$ \\
     $ 15.00$ & $^{\dagger}0.82$ & $^{\dagger}415.$ & $^{\dagger}31.94$ & $   16.133$ & $ 14.68$ & $308.$ & $0.636$ & $0.3742$ & $ 32.4405$ & $  1.3712$ & $    1.211$ & $ 12.33$ & $1.066\cdot 10^{5}$ & $4.979\cdot 10^{6}$ & $0.4316$ & $ 49.7553$ & $  1.6533$ & $    1.599$ & $ 12.17$ & $1.095\cdot 10^{5}$ & $4.690\cdot 10^{8}$ & $0.4707$ & $ 58.6210$ & $  1.9853$ \\
     *$ 15.00$& $0.91$ & $494.$ & $35.19$ & $   15.655$ & $ 14.65$ & $435.$ & $0.897$ & $0.3574$ & $ 39.9355$ & $  1.2728$ & $    1.287$ & $ 13.07$ & $4.753\cdot 10^{4}$ & $2.446\cdot 10^{6}$ & $0.4093$ & $ 57.2083$ & $  1.5153$ & $    0.083$ & $ 12.94$ & $5.451\cdot 10^{4}$ & $2.696\cdot 10^{7}$ & $0.4409$ & $ 64.6910$ & $  1.7625$ \\
\hline
     $ 12.00$ & $0.00$ & $  0.$ & $  0.$ & $   15.310$ & $ 11.95$ & $  0.$ & $0.000$ & $0.2559$ & $  0.2885$ & $  0.1152$ & $    1.976$ & $ 10.95$ & --                  & --                  & $0.2667$ & $  1.5412$ & $  0.3811$ & $    6.465$ & $ 10.88$ & --                  & --                  & $0.2977$ & $  2.4608$ & $  0.5905$ \\
     $ 12.00$ & $^{\dagger}0.43$ & $^{\dagger}186.$ & $^{\dagger}10.81$ & $   18.395$ & $ 11.93$ & $177.$ & $0.400$ & $0.2699$ & $  2.3935$ & $  0.5259$ & $    1.910$ & $ 11.42$ & $2.350\cdot 10^{4}$ & $8.443\cdot 10^{5}$ & $0.3068$ & $  5.3803$ & $  0.8292$ & $    5.168$ & $ 11.34$ & $4.826\cdot 10^{4}$ & $1.910\cdot 10^{8}$ & $0.3499$ & $  7.7164$ & $  1.0572$ \\
     $ 12.00$ & $^{\dagger}0.82$ & $^{\dagger}399.$ & $^{\dagger}20.73$ & $   22.249$ & $ 11.88$ & $378.$ & $0.799$ & $0.3638$ & $ 44.9194$ & $  1.3331$ & $    1.802$ & $ 11.07$ & $2.209\cdot 10^{4}$ & $1.317\cdot 10^{6}$ & $0.4041$ & $ 67.7565$ & $  1.5184$ & $    0.040$ & $ 10.98$ & $3.258\cdot 10^{4}$ & $2.320\cdot 10^{7}$ & $0.4299$ & $ 77.6552$ & $  1.6540$ \\
     *$ 12.00$& $0.91$ & $477.$ & $22.97$ & $   22.297$ & $ 11.84$ & $432.$ & $0.908$ & $0.3577$ & $ 56.0327$ & $  1.3291$ & $    1.877$ & $ 11.29$ & $1.754\cdot 10^{4}$ & $1.045\cdot 10^{6}$ & $0.3975$ & $ 82.0123$ & $  1.5125$ & $    0.075$ & $ 11.19$ & $2.725\cdot 10^{4}$ & $2.078\cdot 10^{7}$ & $0.4221$ & $ 91.6050$ & $  1.6360$ \\
\hline
     $  9.00$ & $0.00$ & $  0.$ & $  0.$ & $   25.992$ & $  8.99$ & $  0.$ & $0.000$ & $0.2559$ & $  0.2885$ & $  0.1152$ & $    3.363$ & $  8.65$ & --                  & --                  & $0.2614$ & $  1.5134$ & $  0.3785$ & $    0.001$ & $  8.61$ & --                  & --                  & $0.2736$ & $  1.8993$ & $  0.4749$ \\
     $  9.00$ & $^{\dagger}0.44$ & $^{\dagger}175.$ & $^{\dagger}6.16$ & $   30.861$ & $  8.99$ & $175.$ & $0.410$ & $0.2649$ & $  1.9041$ & $  0.4538$ & $    3.242$ & $  8.74$ & $1.244\cdot 10^{4}$ & $5.072\cdot 10^{5}$ & $0.3103$ & $  6.2920$ & $  0.8623$ & $    0.047$ & $  8.69$ & $3.279\cdot 10^{4}$ & $5.744\cdot 10^{7}$ & $0.3146$ & $  6.6121$ & $  0.8869$ \\
     $  9.00$ & $^{\dagger}0.82$ & $^{\dagger}377.$ & $^{\dagger}11.79$ & $   35.516$ & $  8.98$ & $394.$ & $0.858$ & $0.3296$ & $ 33.6904$ & $  1.1950$ & $    3.504$ & $  8.90$ & $6.221\cdot 10^{3}$ & $4.743\cdot 10^{5}$ & $0.3678$ & $ 58.1684$ & $  1.3719$ & $    0.057$ & $  8.84$ & $1.436\cdot 10^{4}$ & $3.451\cdot 10^{7}$ & $0.3808$ & $ 67.9385$ & $  1.4335$ \\
     *$  9.00$& $0.92$ & $453.$ & $13.12$ & $   33.818$ & $  8.94$ & $417.$ & $0.899$ & $0.3197$ & $ 40.1316$ & $  1.1893$ & $    3.927$ & $  8.89$ & $5.075\cdot 10^{3}$ & $4.167\cdot 10^{5}$ & $0.3452$ & $ 63.9883$ & $  1.3233$ & $    0.018$ & $  8.84$ & $1.326\cdot 10^{4}$ & $5.316\cdot 10^{7}$ & $0.3581$ & $ 75.0921$ & $  1.3897$ \\
\hline
     $  7.00$ & $0.00$ & $  0.$ & $  0.$ & $   41.395$ & $  7.00$ & $  0.$ & $0.000$ & $0.2559$ & $  0.2885$ & $  0.1152$ & $    6.275$ & $  6.89$ & --                  & --                  & $0.2587$ & $  1.3783$ & $  0.3438$ & & & & & & & \\
     $  7.00$ & $^{\dagger}0.44$ & $^{\dagger}166.$ & $^{\dagger}3.72$ & $   48.956$ & $  7.00$ & $166.$ & $0.407$ & $0.2613$ & $  1.3475$ & $  0.3665$ & $    5.903$ & $  6.88$ & $5.753\cdot 10^{3}$ & $2.729\cdot 10^{5}$ & $0.3020$ & $  5.2802$ & $  0.7972$ & & & & & & & \\
     $  7.00$ & $^{\dagger}0.82$ & $^{\dagger}358.$ & $^{\dagger}7.13$ & $   50.622$ & $  7.00$ & $399.$ & $0.903$ & $0.2898$ & $ 12.0638$ & $  0.9159$ & $    7.729$ & $  6.98$ & $2.594\cdot 10^{3}$ & $2.244\cdot 10^{5}$ & $0.3155$ & $ 23.9778$ & $  1.0888$ & & & & & & & \\
     *$  7.00$& $0.92$ & $432.$ & $7.95$ & $   54.297$ & $  6.96$ & $394.$ & $0.891$ & $0.3080$ & $ 24.5605$ & $  1.0611$ & $    6.748$ & $  6.95$ & $2.140\cdot 10^{3}$ & $1.991\cdot 10^{5}$ & $0.3265$ & $ 38.3670$ & $  1.1710$ & & & & & & & \\
\cline{1-18}
     $  5.00$ & $0.00$ & $  0.$ & $  0.$ & $   83.169$ & $  5.00$ & $  0.$ & $0.000$ & $0.2559$ & $  0.2885$ & $  0.1152$ & $   15.908$ & $  4.95$ & --                  & --                  & $0.2582$ & $  1.3024$ & $  0.3256$ & & & & & & & \\
     $  5.00$ & $^{\dagger}0.44$ & $^{\dagger}157.$ & $^{\dagger}1.87$ & $   99.280$ & $  5.00$ & $152.$ & $0.401$ & $0.2588$ & $  0.8924$ & $  0.2773$ & $   15.334$ & $  4.93$ & $2.292\cdot 10^{3}$ & $1.291\cdot 10^{5}$ & $0.2958$ & $  4.2339$ & $  0.7274$ & & & & & & & \\
     $  5.00$ & $^{\dagger}0.83$ & $^{\dagger}336.$ & $^{\dagger}3.58$ & $  101.879$ & $  5.00$ & $358.$ & $0.888$ & $0.2728$ & $  4.6592$ & $  0.6626$ & $   18.136$ & $  4.95$ & $8.985\cdot 10^{2}$ & $1.009\cdot 10^{5}$ & $0.3070$ & $ 14.3721$ & $  0.9593$ & & & & & & & \\
     *$  5.00$& $0.92$ & $402.$ & $3.98$ & $  108.060$ & $  4.98$ & $374.$ & $0.928$ & $0.2908$ & $ 10.5970$ & $  0.8371$ & $   18.044$ & $  4.92$ & $1.426\cdot 10^{3}$ & $1.527\cdot 10^{5}$ & $0.3203$ & $ 21.9276$ & $  1.0251$ & & & & & & & \\
\cline{1-18}
     $  4.00$ & $0.00$ & $  0.$ & $  0.$ & $  137.775$ & $  4.00$ & $  0.$ & $0.000$ & $0.2559$ & $  0.2885$ & $  0.1152$ & $   33.275$ & $  3.96$ & --                  & --                  & $0.2623$ & $  1.4460$ & $  0.3596$ & & & & & & & \\
     $  4.00$ & $^{\dagger}0.44$ & $^{\dagger}148.$ & $^{\dagger}1.18$ & $  165.639$ & $  4.00$ & $141.$ & $0.396$ & $0.2576$ & $  0.6796$ & $  0.2261$ & $   28.500$ & $  3.95$ & $9.016\cdot 10^{2}$ & $5.691\cdot 10^{4}$ & $0.2970$ & $  3.8659$ & $  0.6970$ & & & & & & & \\
     $  4.00$ & $^{\dagger}0.83$ & $^{\dagger}321.$ & $^{\dagger}2.26$ & $  170.473$ & $  4.00$ & $302.$ & $0.810$ & $0.2692$ & $  3.3039$ & $  0.5652$ & $   35.440$ & $  3.96$ & $4.810\cdot 10^{2}$ & $5.792\cdot 10^{4}$ & $0.3085$ & $ 11.7103$ & $  0.9063$ & & & & & & & \\
     *$  4.00$& $0.94$ & $405.$ & $2.58$ & $  183.355$ & $  3.98$ & $350.$ & $0.951$ & $0.2854$ & $  7.1076$ & $  0.7363$ & $   32.206$ & $  3.93$ & $6.132\cdot 10^{2}$ & $7.238\cdot 10^{4}$ & $0.3222$ & $ 16.8044$ & $  0.9812$ & & & & & & & \\
\cline{1-18}
     $  3.00$ & $0.00$ & $  0.$ & $  0.$ & $  276.224$ & $  3.00$ & $  0.$ & $0.000$ & $0.2559$ & $  0.2885$ & $  0.1152$ & $   91.745$ & $  2.98$ & --                  & --                  & $0.2730$ & $  1.6582$ & $  0.4072$ & & & & & & & \\
     $  3.00$ & $^{\dagger}0.44$ & $^{\dagger}140.$ & $^{\dagger}0.65$ & $  335.387$ & $  3.00$ & $126.$ & $0.394$ & $0.2571$ & $  0.5369$ & $  0.1880$ & $   77.623$ & $  2.97$ & $4.049\cdot 10^{2}$ & $2.891\cdot 10^{4}$ & $0.3054$ & $  3.6971$ & $  0.6851$ & & & & & & & \\
     $  3.00$ & $^{\dagger}0.83$ & $^{\dagger}302.$ & $^{\dagger}1.25$ & $  348.099$ & $  3.00$ & $272.$ & $0.810$ & $0.2668$ & $  2.4237$ & $  0.4789$ & $   83.829$ & $  2.97$ & $1.985\cdot 10^{2}$ & $2.562\cdot 10^{4}$ & $0.3157$ & $  9.5293$ & $  0.8633$ & & & & & & & \\
     *$  3.00$& $0.96$ & $394.$ & $1.45$ & $  377.908$ & $  2.98$ & $276.$ & $0.826$ & $0.2804$ & $  4.7871$ & $  0.6456$ & $   76.037$ & $  2.95$ & $2.959\cdot 10^{2}$ & $3.618\cdot 10^{4}$ & $0.3300$ & $ 13.3044$ & $  0.9758$ & & & & & & & \\
\cline{1-18}
     $  2.50$ & $0.00$ & $  0.$ & $  0.$ & $  440.910$ & $  2.50$ & $  0.$ & $0.000$ & $0.2559$ & $  0.2885$ & $  0.1152$ & $  175.074$ & $  2.48$ & --                  & --                  & $0.2757$ & $  1.6266$ & $  0.3952$ & & & & & & & \\
     $  2.50$ & $^{\dagger}0.44$ & $^{\dagger}134.$ & $^{\dagger}0.45$ & $  541.514$ & $  2.50$ & $121.$ & $0.395$ & $0.2569$ & $  0.4771$ & $  0.1707$ & $  152.438$ & $  2.48$ & $2.381\cdot 10^{2}$ & $1.854\cdot 10^{4}$ & $0.3098$ & $  3.5255$ & $  0.6559$ & & & & & & & \\
     $  2.50$ & $^{\dagger}0.83$ & $^{\dagger}289.$ & $^{\dagger}0.86$ & $  565.360$ & $  2.50$ & $265.$ & $0.831$ & $0.2663$ & $  2.1052$ & $  0.4356$ & $  153.547$ & $  2.48$ & $1.332\cdot 10^{2}$ & $1.914\cdot 10^{4}$ & $0.3190$ & $  8.3689$ & $  0.8175$ & & & & & & & \\
     *$  2.50$& $0.92$ & $345.$ & $0.96$ & $  592.724$ & $  2.49$ & $257.$ & $0.813$ & $0.2741$ & $  3.5273$ & $  0.5284$ & $  138.211$ & $  2.47$ & $1.446\cdot 10^{2}$ & $1.971\cdot 10^{4}$ & $0.3294$ & $ 10.8426$ & $  0.8588$ & & & & & & & \\
\cline{1-18}
     $  2.00$ & $0.00$ & $  0.$ & $  0.$ & $  805.801$ & $  2.00$ & $  0.$ & $0.000$ & $0.2559$ & $  0.2885$ & $  0.1152$ \\
     $  2.00$ & $^{\dagger}0.46$ & $^{\dagger}136.$ & $^{\dagger}0.29$ & $ 1007.382$ & $  2.00$ & $120.$ & $0.412$ & $0.2571$ & $  0.4409$ & $  0.1596$ \\
     $  2.00$ & $^{\dagger}0.84$ & $^{\dagger}278.$ & $^{\dagger}0.56$ & $ 1054.389$ & $  2.00$ & $246.$ & $0.817$ & $0.2671$ & $  1.8486$ & $  0.3842$ \\
     *$  2.00$& $0.94$ & $348.$ & $0.63$ & $ 1118.312$ & $  1.99$ & $243.$ & $0.813$ & $0.2760$ & $  3.1706$ & $  0.4920$\\
\cline{1-11}
     $  1.70$ & $0.00$ & $  0.$ & $  0.$ & $ 1284.504$ & $  1.70$ & $  0.$ & $0.000$ & $0.2559$ & $  0.2885$ & $  0.1152$ \\
     $  1.70$ & $^{\dagger}0.46$ & $^{\dagger}131.$ & $^{\dagger}0.21$ & $ 1616.978$ & $  1.70$ & $117.$ & $0.413$ & $0.2573$ & $  0.4062$ & $  0.1494$ \\
    $  1.70$ & $^{\dagger}0.84$ & $^{\dagger}264.$ & $^{\dagger}0.41$ & $ 1695.775$ & $  1.70$ & $238.$ & $0.812$ & $0.2687$ & $  1.5883$ & $  0.3451$ \\
    *$  1.70$& $0.92$ & $313.$ & $0.46$ & $ 1882.756$ & $  1.69$ & $228.$ & $0.792$ & $0.2822$ & $  3.4912$ & $  0.4830$\\
\hline
\hline
\end{tabular}}
\label{TabListModelsZ006}
\end{sidewaystable*}

\begin{sidewaystable*}
\caption{Main parameters of A-B stars at $Z=0.002=Z_\text{SMC}$}
\centering
\scalebox{0.63}{\begin{tabular}{rccc|ccccccc|ccccccc|ccccccc}
\hline\hline
\multicolumn{4}{c|}{} & \multicolumn{7}{c|}{End of H-burning} & \multicolumn{7}{c|}{End of He-burning} & \multicolumn{7}{c}{End of C-burning}\\
     $M_\text{ini}$ & $\Omega/\Omega_\text{crit, ini}$ & $v_\text{eq}$ & $\mathcal{L}$ & $\tau_\text{H}$ & $M$ & $v_\text{eq}$ & $v_\text{eq}/v_\text{crit}$ & $Y_\text{surf}$ & $\text{N}/\text{C}$ & $\text{N}/\text{O}$ & $\tau_\text{He}$ & $M$ & $P_\text{rot}$ & $\Omega_\text{cen}/\Omega_\text{sur}$ & $Y_\text{surf}$ & $\text{N}/\text{C}$ & $\text{N}/\text{O}$ & $\tau_\text{C}$ & $M$ & $P_\text{rot}$ & $\Omega_\text{cen}/\Omega_\text{sur}$ & $Y_\text{surf}$ & $\text{N}/\text{C}$ & $\text{N}/\text{O}$ \\
$M_{\sun}$ & & km s$^{-1}$& $10^{51}$ g cm$^2$ s$^{-1}$  & Myr & $M_{\sun}$ & km s$^{-1}$ & & \multicolumn{3}{c|}{mass fract.} & Myr & $M_{\sun}$ & day & & \multicolumn{3}{c|}{mass fract.} & kyr & $M_{\sun}$ & day & & \multicolumn{3}{c}{mass fract.}\\
\hline
     $ 15.00$ & $0.00$ & $  0.$ & $  0.$ & $   10.996$ & $ 14.92$ & $  0.$ & $0.000$ & $0.2509$ & $  0.2885$ & $  0.1152$ & $    1.363$ & $ 14.69$ & --                  & --                  & $0.2538$ & $  1.0804$ & $  0.2851$ & $    2.143$ & $ 14.60$ & --                  & --                  & $0.3078$ & $  2.6821$ & $  0.6375$ \\
     $ 15.00$ & $^{\dagger}0.43$ & $^{\dagger}207.$ & $^{\dagger}17.06$ & $   13.174$ & $ 14.88$ & $191.$ & $0.391$ & $0.2757$ & $  4.5795$ & $  0.7518$ & $    1.323$ & $ 14.48$ & $2.390\cdot 10^{4}$ & $7.992\cdot 10^{5}$ & $0.3104$ & $  8.3919$ & $  1.0121$ & $    2.398$ & $ 14.35$ & $3.577\cdot 10^{4}$ & $1.037\cdot 10^{8}$ & $0.3762$ & $ 13.5690$ & $  1.5163$ \\
     $ 15.00$ & $^{\dagger}0.82$ & $447.$ & $^{\dagger}32.82$ & $   16.820$ & $ 14.77$ & $363.$ & $0.699$ & $0.4126$ & $105.3703$ & $  1.8105$ & $    1.118$ & $ 14.24$ & $2.044\cdot 10^{4}$ & $1.026\cdot 10^{6}$ & $0.4436$ & $128.1636$ & $  1.9873$ & $    1.136$ & $ 14.05$ & $2.966\cdot 10^{4}$ & $1.244\cdot 10^{8}$ & $0.4685$ & $133.5483$ & $  2.1527$ \\
     *$ 15.00$& $0.92$& $538.$ & $37.15$ & $   16.794$ & $ 14.71$ & $410.$ & $0.799$ & $0.3961$ & $129.2907$ & $  1.7362$ & $    1.143$ & $ 13.92$ & $2.723\cdot 10^{4}$ & $1.337\cdot 10^{6}$ & $0.4215$ & $147.1358$ & $  1.8705$ & $    0.048$ & $ 13.75$ & $3.255\cdot 10^{4}$ & $1.351\cdot 10^{7}$ & $0.4612$ & $141.9440$ & $  2.2448$\\
\hline
     $ 12.00$ & $0.00$ & $  0.$ & $  0.$ & $   15.100$ & $ 11.97$ & $  0.$ & $0.000$ & $0.2509$ & $  0.2885$ & $  0.1152$ & $    1.852$ & $ 11.91$ & --                  & --                  & $0.2509$ & $  0.9303$ & $  0.2531$ & $    6.643$ & $ 11.83$ & --                  & --                  & $0.2900$ & $  2.4466$ & $  0.5674$ \\
     $ 12.00$ & $^{\dagger}0.43$ & $^{\dagger}198.$ & $^{\dagger}11.10$ & $   18.006$ & $ 11.95$ & $188.$ & $0.395$ & $0.2706$ & $  4.4735$ & $  0.7233$ & $    1.841$ & $ 11.89$ & $5.165\cdot 10^{3}$ & $1.925\cdot 10^{5}$ & $0.2733$ & $  4.8518$ & $  0.7520$ & $    6.206$ & $ 11.80$ & $2.328\cdot 10^{4}$ & $1.093\cdot 10^{8}$ & $0.3488$ & $ 12.9178$ & $  1.2772$ \\
    $ 12.00$ & $^{\dagger}0.82$ & $^{\dagger}423.$ & $^{\dagger}21.37$ & $   22.266$ & $ 11.91$ & $426.$ & $0.837$ & $0.3957$ & $132.0904$ & $  1.7541$ & $    1.645$ & $ 11.81$ & $1.157\cdot 10^{3}$ & $7.235\cdot 10^{4}$ & $0.4008$ & $136.4635$ & $  1.7816$ & $    0.068$ & $ 11.71$ & $1.564\cdot 10^{4}$ & $1.081\cdot 10^{7}$ & $0.4392$ & $164.1779$ & $  2.0035$ \\
    *$ 12.00$& $0.93$ & $523.$ & $24.46$ & $   22.763$ & $ 11.87$ & $443.$ & $0.888$ & $0.3690$ & $156.8571$ & $  1.6748$ & $    1.759$ & $ 11.78$ & $2.682\cdot 10^{3}$ & $1.611\cdot 10^{5}$ & $0.3735$ & $161.4954$ & $  1.6988$ & $    0.072$ & $ 11.68$ & $1.491\cdot 10^{4}$ & $1.117\cdot 10^{7}$ & $0.4114$ & $180.3932$ & $  1.9223$ \\

\hline
     $  9.00$ & $0.00$ & $  0.$ & $  0.$ & $   25.471$ & $  8.99$ & $  0.$ & $0.000$ & $0.2509$ & $  0.2885$ & $  0.1152$ & $    2.999$ & $  8.89$ & --                  & --                  & $0.2510$ & $  1.2200$ & $  0.2935$ & $    0.047$ & $  8.85$ & --                  & --                  & $0.2623$ & $  2.0719$ & $  0.4711$ \\
     $  9.00$ & $^{\dagger}0.43$ & $^{\dagger}185.$ & $^{\dagger}6.35$ & $   30.090$ & $  8.99$ & $180.$ & $0.392$ & $0.2659$ & $  3.9935$ & $  0.6728$ & $    3.170$ & $  8.97$ & $2.075\cdot 10^{3}$ & $8.907\cdot 10^{4}$ & $0.2666$ & $  4.1190$ & $  0.6828$ & $    0.061$ & $  8.92$ & $1.941\cdot 10^{4}$ & $2.626\cdot 10^{7}$ & $0.3065$ & $ 10.6774$ & $  1.0264$ \\
     $  9.00$ & $^{\dagger}0.82$ & $^{\dagger}400.$ & $^{\dagger}12.22$ & $   37.256$ & $  8.98$ & $430.$ & $0.878$ & $0.3804$ & $146.9191$ & $  1.7473$ & $    2.821$ & $  8.95$ & $1.761\cdot 10^{2}$ & $1.356\cdot 10^{4}$ & $0.3838$ & $150.6317$ & $  1.7659$ & $    0.076$ & $  8.89$ & $1.064\cdot 10^{4}$ & $2.167\cdot 10^{7}$ & $0.4219$ & $187.4874$ & $  1.9825$ \\
     *$  9.00$& $0.93$& $490.$ & $13.91$ & $   37.818$ & $  8.95$ & $434.$ & $0.882$ & $0.3777$ & $190.4415$ & $  1.7518$ & $    2.969$ & $  8.93$ & $1.132\cdot 10^{2}$ & $9.112\cdot 10^{3}$ & $0.3800$ & $192.4962$ & $  1.7642$ & $    0.066$ & $  8.87$ & $1.009\cdot 10^{4}$ & $2.379\cdot 10^{7}$ & $0.4113$ & $213.4628$ & $  1.9412$ \\
\hline
     $  7.00$ & $0.00$ & $  0.$ & $  0.$ & $   39.945$ & $  7.00$ & $  0.$ & $0.000$ & $0.2509$ & $  0.2885$ & $  0.1152$ & $    5.570$ & $  6.94$ & --                  & --                  & $0.2511$ & $  1.0737$ & $  0.2745$ & & & & & & & \\
     $  7.00$ & $^{\dagger}0.43$ & $^{\dagger}177.$ & $^{\dagger}3.86$ & $   46.821$ & $  7.00$ & $171.$ & $0.388$ & $0.2615$ & $  3.0065$ & $  0.5893$ & $    5.565$ & $  6.94$ & $3.677\cdot 10^{2}$ & $1.780\cdot 10^{4}$ & $0.2834$ & $  6.4559$ & $  0.8439$ & & & & & & & \\
     $  7.00$ & $^{\dagger}0.82$ & $^{\dagger}379.$ & $^{\dagger}7.42$ & $   56.509$ & $  7.00$ & $413.$ & $0.879$ & $0.3450$ & $ 93.7738$ & $  1.5255$ & $    5.101$ & $  6.99$ & $1.665\cdot 10^{2}$ & $1.608\cdot 10^{4}$ & $0.3485$ & $ 98.0867$ & $  1.5439$ & & & & & & & \\
     *$  7.00$& $0.93$& $461.$ & $8.20$ & $   63.425$ & $  6.97$ & $428.$ & $0.886$ & $0.4209$ & $227.4597$ & $  2.0013$ & $    4.153$ & $  6.95$ & $7.462\cdot 10^{1}$ & $7.133\cdot 10^{3}$ & $0.4235$ & $228.7656$ & $  2.7656$ & & & & & & & \\
\cline{1-18}
     $  5.00$ & $0.00$ & $  0.$ & $  0.$ & $   77.356$ & $  5.00$ & $  0.$ & $0.000$ & $0.2509$ & $  0.2885$ & $  0.1152$ & $   13.652$ & $  4.96$ & --                  & --                  & $0.2513$ & $  1.0048$ & $  0.2649$ & & & & & & & \\
     $  5.00$ & $^{\dagger}0.43$ & $^{\dagger}167.$ & $^{\dagger}1.97$ & $   91.541$ & $  5.00$ & $159.$ & $0.384$ & $0.2576$ & $  1.9886$ & $  0.4722$ & $   12.905$ & $  4.95$ & $1.789\cdot 10^{3}$ & $1.034\cdot 10^{5}$ & $0.2793$ & $  5.0756$ & $  0.7719$ & & & & & & & \\
     $  5.00$ & $^{\dagger}0.82$ & $^{\dagger}355.$ & $^{\dagger}3.78$ & $   95.799$ & $  5.00$ & $378.$ & $0.853$ & $0.2899$ & $ 24.3400$ & $  1.1364$ & $   14.362$ & $  5.00$ & $2.397\cdot 10^{2}$ & $2.973\cdot 10^{4}$ & $0.2925$ & $ 26.5232$ & $  1.1567$ & & & & & & & \\
     *$  5.00$& $0.92$& $435.$ & $4.24$ & $  121.500$ & $  5.00$ & $386.$ & $0.858$ & $0.3933$ & $180.5181$ & $  1.7117$ & $    9.306$ & $  4.99$ & $4.268\cdot 10^{2}$ & $5.146\cdot 10^{4}$ & $0.3986$ & $187.5579$ & $  1.7411$ & & & & & & & \\
\cline{1-18}
     $  4.00$ & $0.00$ & $  0.$ & $  0.$ & $  124.375$ & $  4.00$ & $  0.$ & $0.000$ & $0.2509$ & $  0.2885$ & $  0.1152$ & $   26.143$ & $  3.97$ & --                  & --                  & $0.2519$ & $  1.1545$ & $  0.2856$ & & & & & & & \\
     $  4.00$ & $^{\dagger}0.435$ & $^{\dagger}158.$ & $^{\dagger}1.25$ & $  148.207$ & $  4.00$ & $150.$ & $0.383$ & $0.2557$ & $  1.5026$ & $  0.3932$ & $   22.925$ & $  3.96$ & $1.046\cdot 10^{3}$ & $6.835\cdot 10^{4}$ & $0.2793$ & $  4.5882$ & $  0.7309$ & & & & & & & \\
     $  4.00$ & $^{\dagger}0.82$ & $^{\dagger}340.$ & $^{\dagger}2.40$ & $  152.800$ & $  4.00$ & $376.$ & $0.895$ & $0.2812$ & $ 14.8199$ & $  0.9692$ & $   27.807$ & $  3.96$ & $5.956\cdot 10^{2}$ & $7.748\cdot 10^{4}$ & $0.2926$ & $ 21.8880$ & $  1.0603$ & & & & & & & \\
     *$  4.00$& $0.93$& $424.$ & $2.69$  & $  172.360$ & $  3.99$ & $350.$ & $0.825$ & $0.3196$ & $ 56.9703$ & $  1.2015$ & $   24.897$ & $  3.94$ & $9.272\cdot 10^{2}$ & $1.068\cdot 10^{5}$ & $0.3394$ & $ 85.3472$ & $  1.2996$ & & & & & & & \\
\cline{1-18}
     $  3.00$ & $0.00$ & $  0.$ & $  0.$ & $  240.308$ & $  3.00$ & $  0.$ & $0.000$ & $0.2509$ & $  0.2885$ & $  0.1152$ & $   63.990$ & $  2.97$ & --                  & --                  & $0.2549$ & $  1.4165$ & $  0.3226$ & & & & & & & \\
     $  3.00$ & $^{\dagger}0.44$ & $^{\dagger}150.$ & $^{\dagger}0.70$ & $  288.299$ & $  3.00$ & $139.$ & $0.386$ & $0.2546$ & $  1.1508$ & $  0.3291$ & $   54.555$ & $  2.96$ & $5.277\cdot 10^{2}$ & $3.858\cdot 10^{4}$ & $0.2867$ & $  4.9278$ & $  0.7500$ & & & & & & & \\
     $  3.00$ & $^{\dagger}0.82$ & $^{\dagger}318.$ & $^{\dagger}1.34$ & $  296.847$ & $  3.00$ & $313.$ & $0.813$ & $0.2715$ & $  7.6469$ & $  0.7734$ & $   65.639$ & $  2.96$ & $2.990\cdot 10^{2}$ & $4.375\cdot 10^{4}$ & $0.3018$ & $ 21.9796$ & $  1.0155$ & & & & & & & \\
     *$  3.00$ & $0.93$& $394.$ & $1.48$  & $  304.160$ & $  2.99$ & $310.$ & $0.789$ & $0.2890$ & $ 19.4074$ & $  0.9018$ & $   68.181$ & $  2.94$ & $4.295\cdot 10^{2}$ & $5.983\cdot 10^{4}$ & $0.3106$ & $ 36.2952$ & $  1.0292$ & & & & & & & \\
\cline{1-18}
     $  2.50$ & $0.00$ & $  0.$ & $  0.$ & $  374.191$ & $  2.50$ & $  0.$ & $0.000$ & $0.2509$ & $  0.2885$ & $  0.1152$ & $  114.955$ & $  2.48$ & --                  & --                  & $0.2648$ & $  1.7337$ & $  0.3915$ & & & & & & & \\
     $  2.50$ & $^{\dagger}0.43$ & $^{\dagger}147.$ & $^{\dagger}0.49$ & $  453.569$ & $  2.50$ & $133.$ & $0.389$ & $0.2542$ & $  0.9369$ & $  0.2794$ & $   97.722$ & $  2.47$ & $3.093\cdot 10^{2}$ & $2.553\cdot 10^{4}$ & $0.2962$ & $  4.9999$ & $  0.7310$ & & & & & & & \\
     $  2.50$ & $^{\dagger}0.82$ & $^{\dagger}306.$ & $^{\dagger}0.93$ & $  468.888$ & $  2.50$ & $311.$ & $0.841$ & $0.2695$ & $  5.7329$ & $  0.6636$ & $  112.967$ & $  2.47$ & $1.663\cdot 10^{2}$ & $2.674\cdot 10^{4}$ & $0.3095$ & $ 22.3530$ & $  0.9596$ & & & & & & & \\
     *$  2.50$& $0.93$ & $389.$ & $1.10$  & $  489.630$ & $  2.49$ & $312.$ & $0.839$ & $0.2837$ & $ 12.8815$ & $  0.8177$ & $  117.680$ & $  2.45$ & $2.375\cdot 10^{2}$ & $3.423\cdot 10^{4}$ & $0.3210$ & $ 36.8810$ & $  1.0411$ & & & & & & & \\
\cline{1-18}
     $  2.00$ & $0.00$ & $  0.$ & $  0.$ & $  667.029$ & $  2.00$ & $  0.$ & $0.000$ & $0.2509$ & $  0.2885$ & $  0.1152$ & & & & & & & & & & & & & & \\
     $  2.00$ & $^{\dagger}0.44$ & $^{\dagger}138.$ & $^{\dagger}0.32$ & $  825.594$ & $  2.00$ & $134.$ & $0.410$ & $0.2547$ & $  0.8046$ & $  0.2451$ & $  194.067$ & $  1.97$ & $2.192\cdot 10^{2}$ & $2.047\cdot 10^{4}$ & $0.3032$ & $  4.7208$ & $  0.6659$ & & & & & & & \\
     $  2.00$ & $^{\dagger}0.82$ & $^{\dagger}290.$ & $^{\dagger}0.61$ & $  871.563$ & $  2.00$ & $298.$ & $0.842$ & $0.2753$ & $  5.6638$ & $  0.6399$ & & & & & & & & & & & & & &  \\
     *$  2.00$& $0.94$ & $375.$ & $0.71$  & $  941.300$ & $  1.99$ & $296.$ & $0.834$ & $0.2904$ & $ 12.5988$ & $  0.8781$ & & & & & & & & & & & & & &  \\
\cline{1-18}
     $  1.70$ & $0.00$ & $  0.$ & $  0.$ & $ 1053.592$ & $  1.70$ & $  0.$ & $0.000$ & $0.2509$ & $  0.2885$ & $  0.1152$ \\
     $  1.70$ & $^{\dagger}0.44$ & $^{\dagger}132.$ & $^{\dagger}0.23$ & $ 1313.265$ & $  1.70$ & $134.$ & $0.421$ & $0.2552$ & $  0.6981$ & $  0.2181$ \\
     $  1.70$ & $^{\dagger}0.81$ & $^{\dagger}265.$ & $^{\dagger}0.45$ & $ 1701.731$ & $  1.70$ & $272.$ & $0.798$ & $0.3111$ & $ 12.7300$ & $  0.9124$ \\
     *$  1.70$& $0.94$ & $352.$ & $0.52$   & $ 1609.072$ & $  1.69$ & $288.$ & $0.850$ & $0.3061$ & $ 15.9100$ & $  0.8343$ \\ 
\hline
\hline
\end{tabular}}
\label{TabListModelsZ002}
\end{sidewaystable*}
\begin{table*}
\caption{Data for the models that reach the critical velocity during the MS.}
\centering
\scalebox{0.7}{\begin{tabular}{rrc||ccccccc}
\hline\hline
\rule[0mm]{0mm}{3mm}$Z$ & $M_\text{ini}$ & $\Omega/\Omega_\text{crit, ini}$ & $\tau_\text{crit. rot.}$ & $\tau_\text{crit. rot.}/\tau_\text{MS}$ & $H_\text{cen}$ & $\Delta M_\text{tot}$ & $\Delta M_\text{rad}$ & $\Delta M_\text{mech}$ & $\dot{M}_\text{mech, mean}$ \\ 
\rule[-1.5mm]{0mm}{2mm} & $M_{\sun}$ & & $\text{yr}$ & & & $M_{\sun}$ & $M_{\sun}$ & $M_{\sun}$ & $M_{\sun}\cdot \text{yr}^{-1}$\\
\hline
\rule[0mm]{0mm}{3mm}
    $ 0.014$& $  1.70$ & $0.95$ & $1.9195\cdot 10^{9}$ & $0.804$ & $0.643$ & $0.0979$ & $0.0892$ & $8.6831\cdot 10^{-3}$ & $4.5236\cdot 10^{-12}$\\
    $ 0.014$& $  2.00$ & $0.96$ & $1.1401\cdot 10^{9}$ & $0.809$ & $0.619$ & $0.0399$ & $0.0310$ & $8.9237\cdot 10^{-3}$ & $7.8274\cdot 10^{-12}$\\
    $ 0.014$& $  2.50$ & $0.95$ & $5.3951\cdot 10^{8}$ & $0.733$ & $0.584$ & $0.0760$ & $0.0651$ & $1.0933\cdot 10^{-2}$ & $2.0265\cdot 10^{-11}$\\
    $ 0.014$& $  3.00$ & $0.95$ & $3.5548\cdot 10^{8}$ & $0.805$ & $0.622$ & $0.0813$ & $0.0641$ & $1.7220\cdot 10^{-2}$ & $4.8443\cdot 10^{-11}$\\
    $ 0.014$& $  4.00$ & $0.95$ & $1.7809\cdot 10^{8}$ & $0.867$ & $0.654$ & $0.1238$ & $0.0990$ & $2.4704\cdot 10^{-2}$ & $1.3872\cdot 10^{-10}$\\
    $ 0.014$& $  5.00$ & $0.95$ & $8.4680\cdot 10^{7}$ & $0.721$ & $0.593$ & $0.1037$ & $0.0756$ & $2.8118\cdot 10^{-2}$ & $3.3205\cdot 10^{-10}$\\
    $ 0.014$& $  7.00$ & $0.95$ & $3.7185\cdot 10^{7}$ & $0.685$ & $0.576$ & $0.1537$ & $0.1159$ & $3.7842\cdot 10^{-2}$ & $1.0177\cdot 10^{-9}$\\
    $ 0.014$& $  9.00$ & $0.95$ & $2.2899\cdot 10^{7}$ & $0.690$ & $0.578$ & $0.2289$ & $0.1825$ & $4.6381\cdot 10^{-2}$ & $2.0255\cdot 10^{-9}$\\
    $ 0.014$& $ 12.00$ & $0.95$ & $1.1726\cdot 10^{7}$ & $0.541$ & $0.589$ & $1.5632$ & $1.5489$ & $1.4287\cdot 10^{-2}$ & $1.2184\cdot 10^{-9}$\\
    $ 0.014$& $ 15.00$ & $0.95$ & $9.1153\cdot 10^{6}$ & $0.582$ & $0.574$ & $3.1693$ & $3.1637$ & $5.6247\cdot 10^{-3}$ & $6.1706\cdot 10^{-10}$\\\hline
    $ 0.006$& $  1.70$ & $0.93$ & $1.3225\cdot 10^{9}$ & $0.702$ & $0.596$ & $0.0818$ & $0.0719$ & $9.8766\cdot 10^{-3}$ & $7.4683\cdot 10^{-12}$\\
    $ 0.006$& $  2.00$ & $0.94$ & $8.9484\cdot 10^{8}$ & $0.800$ & $0.636$ & $0.0361$ & $0.0245$ & $1.1598\cdot 10^{-2}$ & $1.2960\cdot 10^{-11}$\\
    $ 0.006$& $  2.50$ & $0.92$ & $3.8193\cdot 10^{8}$ & $0.644$ & $0.556$ & $0.0385$ & $0.0269$ & $1.1644\cdot 10^{-2}$ & $3.0486\cdot 10^{-11}$\\
    $ 0.006$& $  3.00$ & $0.96$ & $3.1073\cdot 10^{8}$ & $0.822$ & $0.655$ & $0.0575$ & $0.0399$ & $1.7599\cdot 10^{-2}$ & $5.6638\cdot 10^{-11}$\\
    $ 0.006$ & $  4.00$ & $0.94$ & $1.2853\cdot 10^{8}$ & $0.701$ & $0.600$ & $0.0775$ & $0.0580$ & $1.9500\cdot 10^{-2}$ & $1.5172\cdot 10^{-10}$\\
    $ 0.006$& $  5.00$ & $0.92$ & $5.5128\cdot 10^{7}$ & $0.510$ & $0.500$ & $0.0808$ & $0.0630$ & $1.7803\cdot 10^{-2}$ & $3.2294\cdot 10^{-10}$\\
    $ 0.006$& $  7.00$ & $0.92$ & $2.8344\cdot 10^{7}$ & $0.522$ & $0.510$ & $0.7640$ & $0.0508$ & $2.5565\cdot 10^{-2}$ & $9.0195\cdot 10^{-10}$\\
    $ 0.006$& $  9.00$ & $0.92$ & $1.6014\cdot 10^{7}$ & $0.474$ & $0.478$ & $0.1504$ & $0.1235$ & $2.6973\cdot 10^{-2}$ & $1.6843\cdot 10^{-9}$\\
    $ 0.006$ & $ 12.00$ & $0.91$ & $1.1425\cdot 10^{7}$ & $0.512$ & $0.460$ & $0.8014$ & $0.7998$ & $1.6202\cdot 10^{-3}$ & $1.4181\cdot 10^{-10}$\\
    $ 0.006$& $ 15.00$ & $0.91$ & $6.6927\cdot 10^{6}$ & $0.428$ & $0.395$ & $2.0345$ & $2.0328$ & $1.7458\cdot 10^{-3}$ & $2.6084\cdot 10^{-10}$\\\hline
    $ 0.002$& $  1.70$ & $0.94$ & $1.1390\cdot 10^{9}$ & $0.708$ & $0.667$ & $0.0668$ & $0.0556$ & $1.1169\cdot 10^{-2}$ & $9.8063\cdot 10^{-12}$\\
    $ 0.002$& $  2.00$ & $0.94$ & $7.2368\cdot 10^{8}$ & $0.769$ & $0.668$ & $0.0356$ & $0.0256$ & $1.0052\cdot 10^{-2}$ & $1.3890\cdot 10^{-11}$\\
    $ 0.002$& $  2.50$ & $0.93$ & $3.4795\cdot 10^{8}$ & $0.711$ & $0.616$ & $0.0836$ & $0.0736$ & $9.9490\cdot 10^{-3}$ & $2.8593\cdot 10^{-11}$\\
    $ 0.002$& $  3.00$ & $0.93$ & $1.6575\cdot 10^{8}$ & $0.545$ & $0.540$ & $0.0813$ & $0.0715$ & $9.8707\cdot 10^{-3}$ & $5.9551\cdot 10^{-11}$\\
    $ 0.002$& $  4.00$ & $0.93$ & $9.0014\cdot 10^{7}$ & $0.522$ & $0.505$ & $0.0876$ & $0.0756$ & $1.1992\cdot 10^{-2}$ & $1.3322\cdot 10^{-10}$\\
    $ 0.002$& $  5.00$ & $0.92$ & $5.1098\cdot 10^{7}$ & $0.421$ & $0.366$ & $0.0510$ & $0.0412$ & $9.7813\cdot 10^{-3}$ & $1.9142\cdot 10^{-10}$ \\
    $ 0.002$& $  7.00$ & $0.93$ & $1.9022\cdot 10^{7}$ & $0.300$ & $0.235$ & $0.0626$ & $0.0451$ & $1.7543\cdot 10^{-2}$ & $9.2228\cdot 10^{-10}$\\ 
    $ 0.002$& $  9.00$ & $0.93$ & $2.3485\cdot 10^{7}$ & $0.621$ & $0.562$ & $0.1365$ & $0.1042$ & $3.2377\cdot 10^{-2}$ & $1.3786\cdot 10^{-9}$\\
    $ 0.002$& $ 12.00$ & $0.93$ & $1.5147\cdot 10^{7}$ & $0.665$ & $0.576$ & $0.3303$ & $0.2946$ & $3.5740\cdot 10^{-2}$ & $2.3595\cdot 10^{-9}$\\
    $ 0.002$& $ 15.00$ & $0.92$ & $1.0951\cdot 10^{7}$ & $0.652$ & $0.547$ & $1.2601$ & $1.1999$ & $6.0194\cdot 10^{-2}$ & $5.4964\cdot 10^{-9}$\\\hline
\end{tabular}}
\tablefoot{Initial metallicity, mass, and rotation rate (columns 1 to 3), time spent at the critical limit in Myr (column 4) and in units of the MS lifetime (column 5), central hydrogen mass fraction when the critical limit is reached for the first time (column 6), total mass lost during the whole computed evolution (column 7), mass lost through stellar winds (column 8), mass lost mechanically at the critical limit (column 9), mechanical mass-loss rate averaged over the critical-rotation period (column 10).}
\label{TabBeModels}
\end{table*}
\subsection{Stellar lifetimes, mass of the convective cores, and evolutionary tracks in the HRD}

\begin{figure*}
\centerline{\includegraphics[width=0.24\textwidth, angle=270]{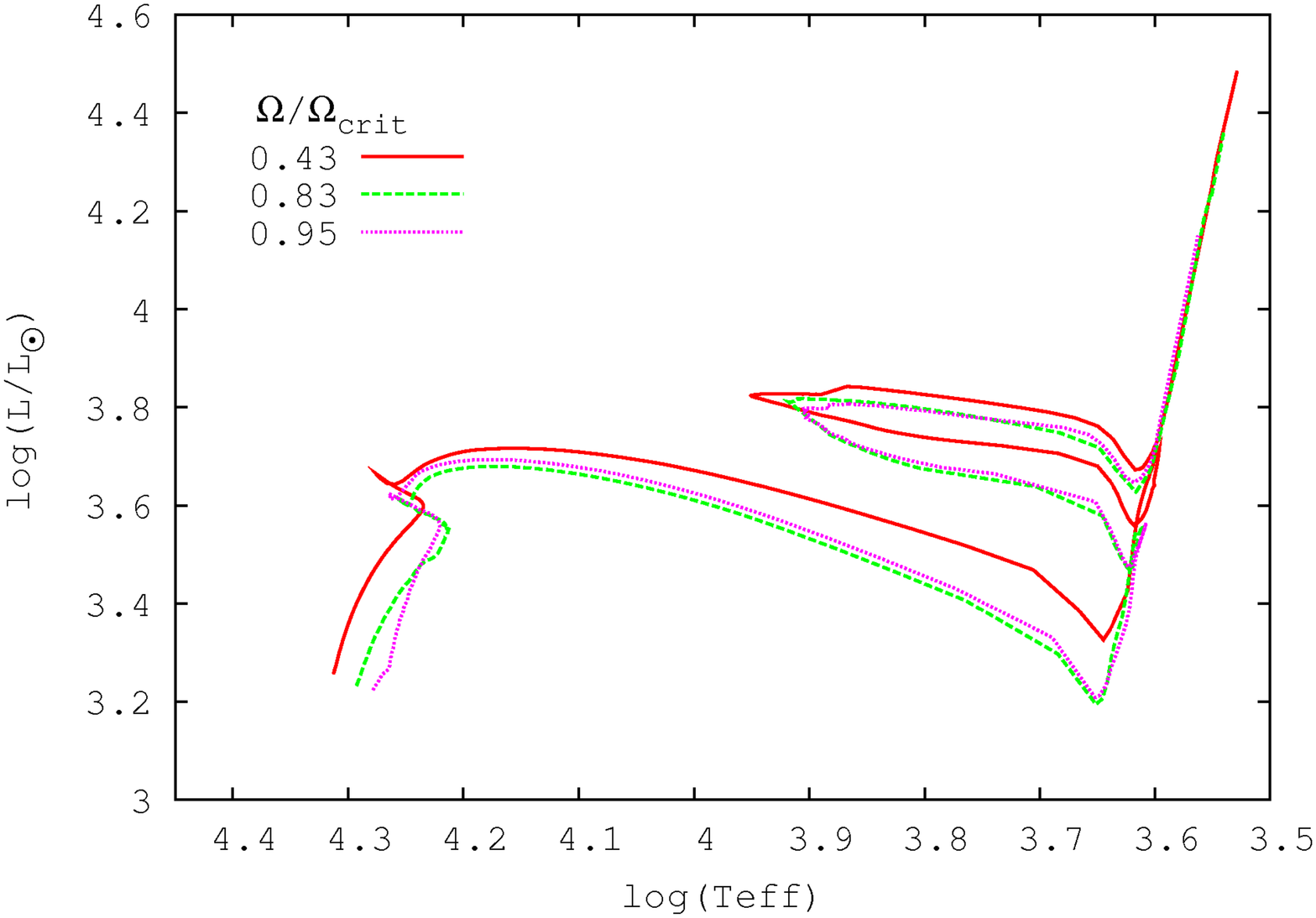}
\includegraphics[width=0.24\textwidth, angle=270]{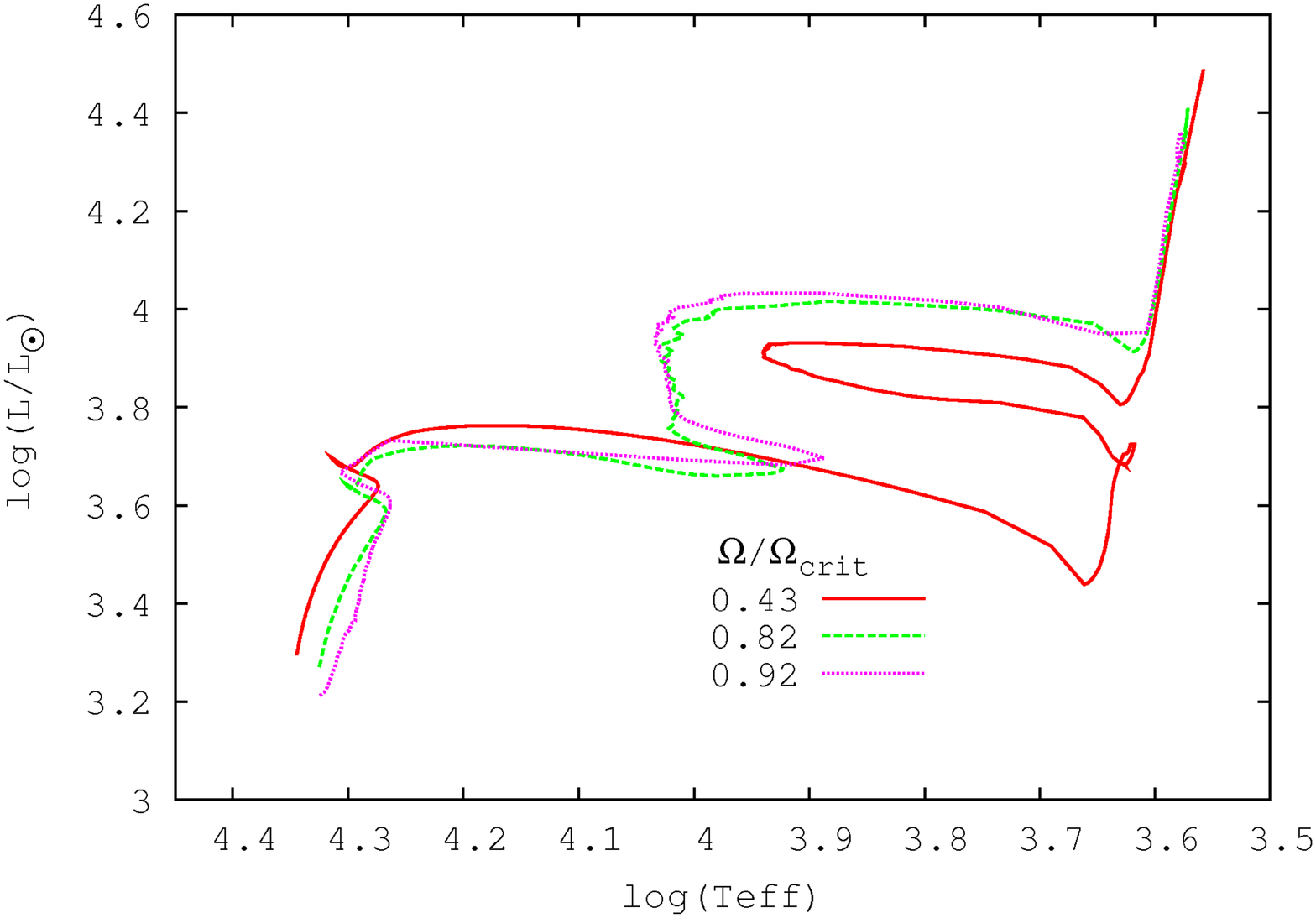}
\includegraphics[width=0.24\textwidth, angle=270]{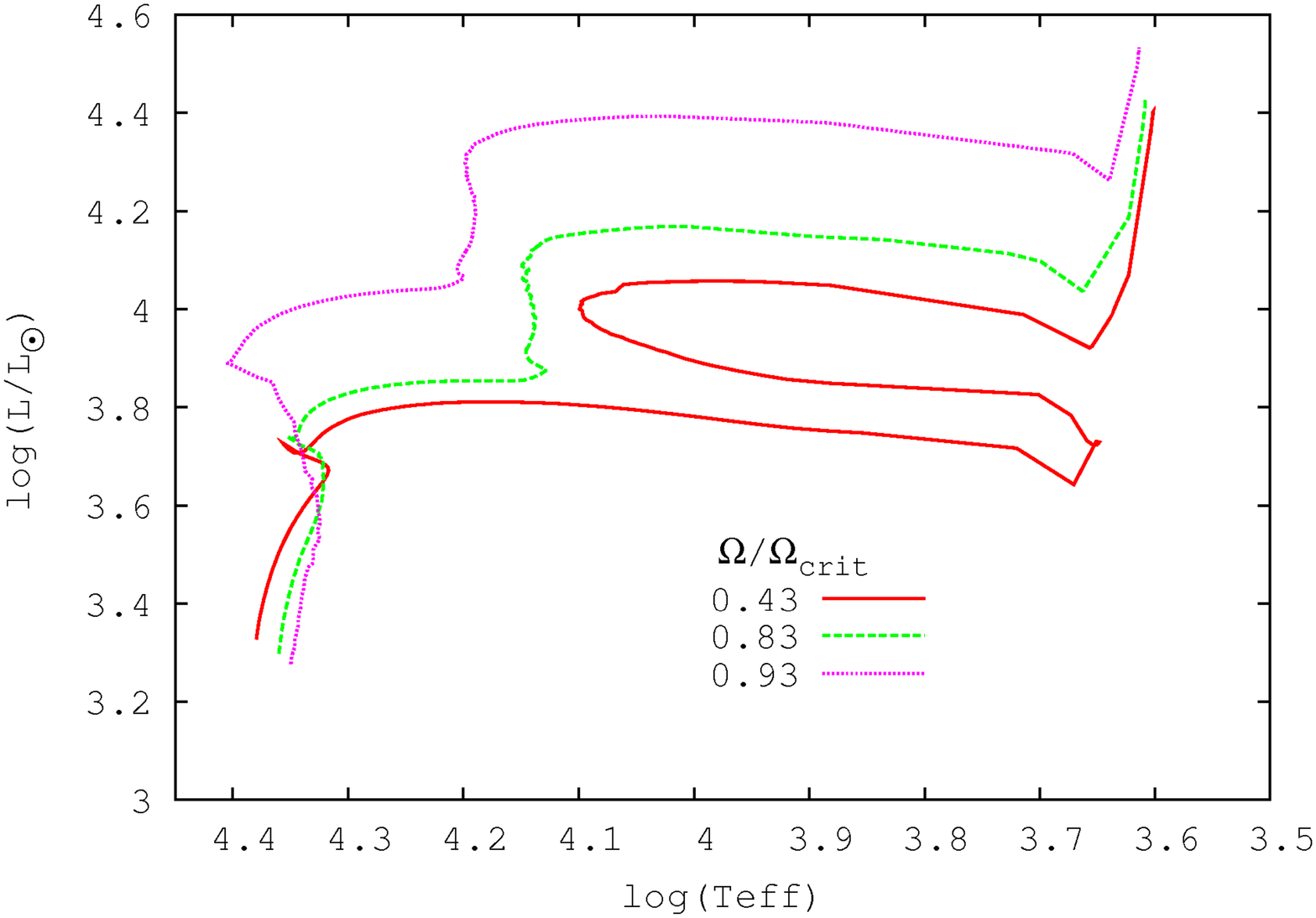}}
\centerline{\includegraphics[width=0.24\textwidth, angle=270]{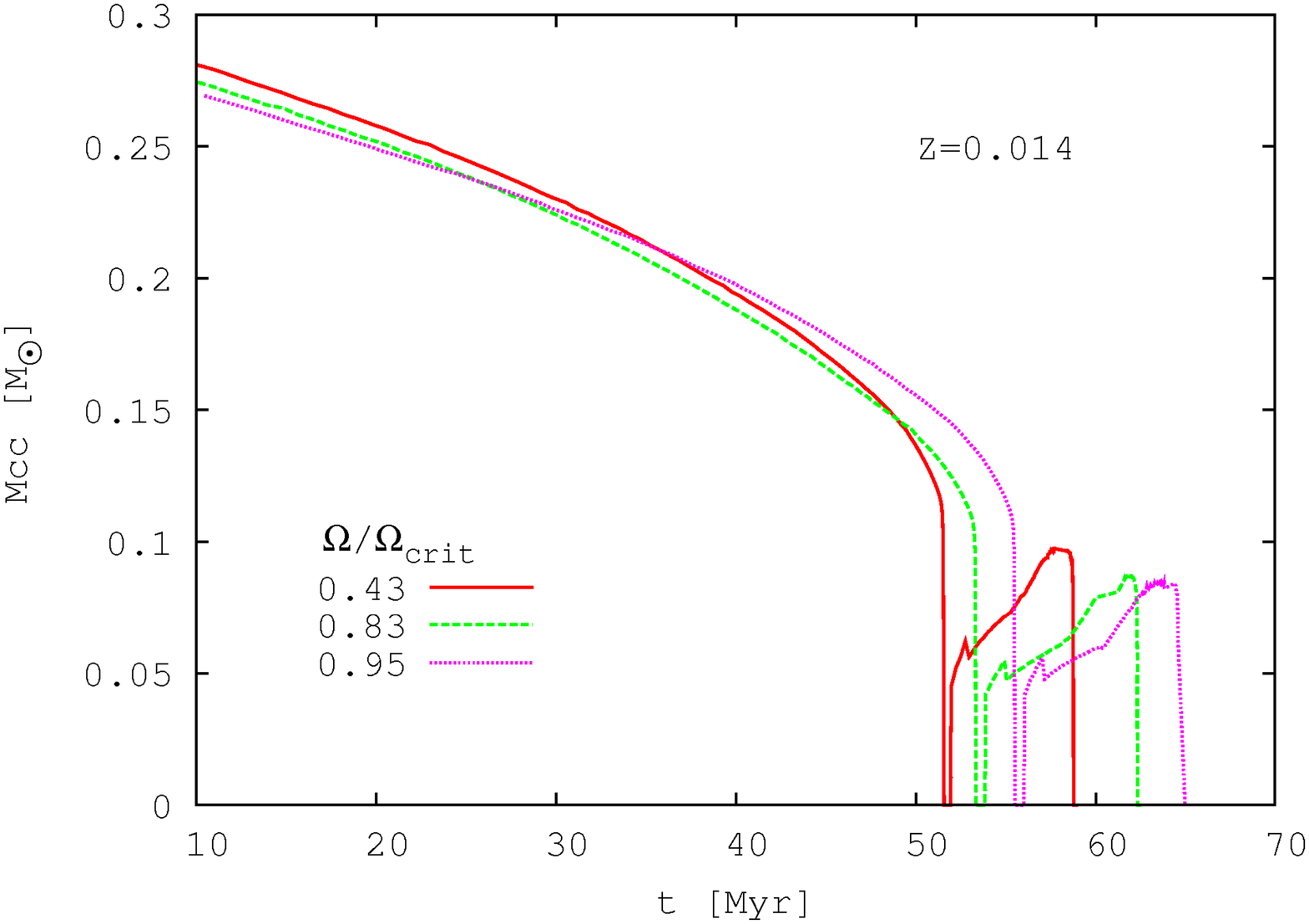}
\includegraphics[width=0.24\textwidth, angle=270]{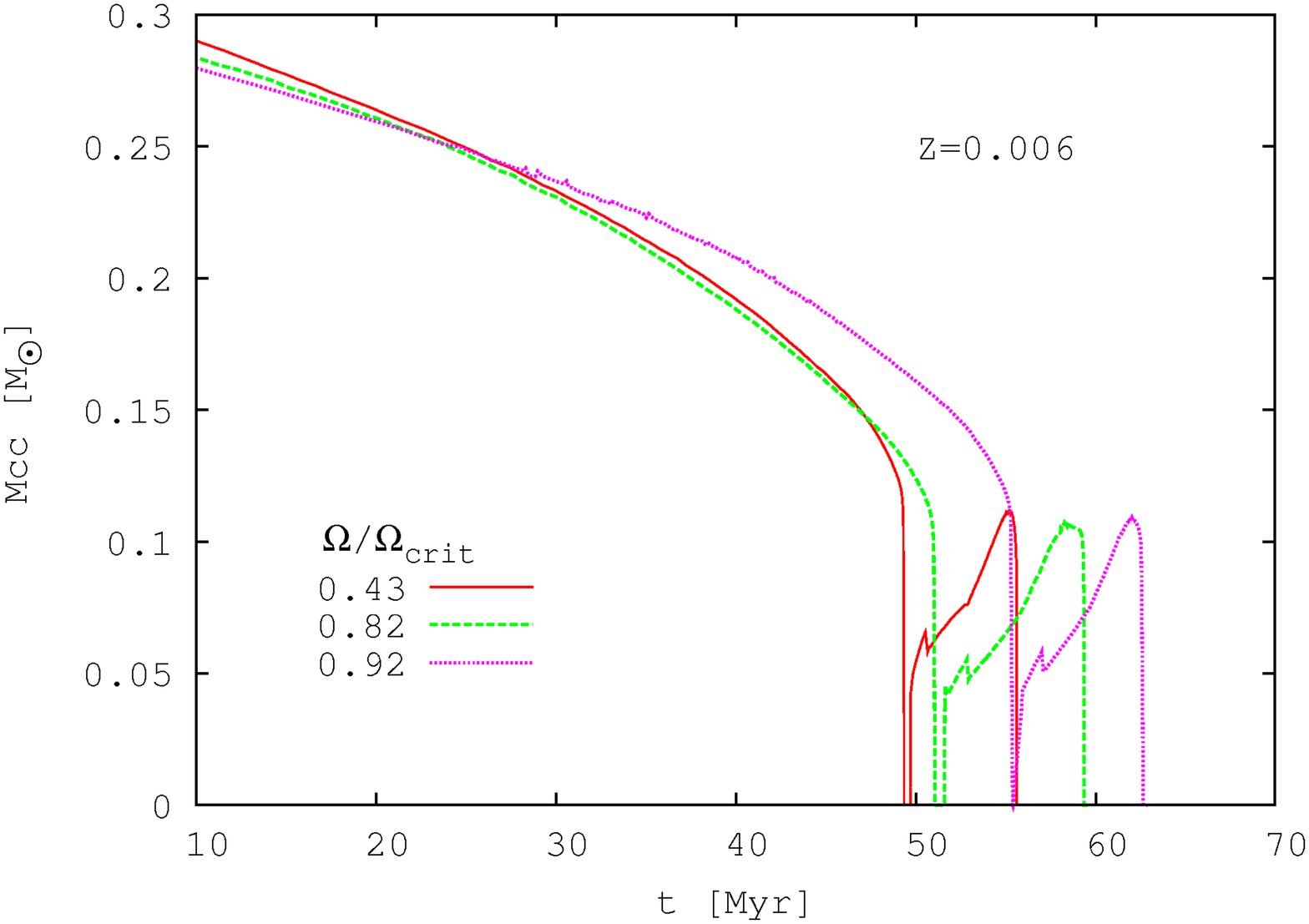}
\includegraphics[width=0.24\textwidth, angle=270]{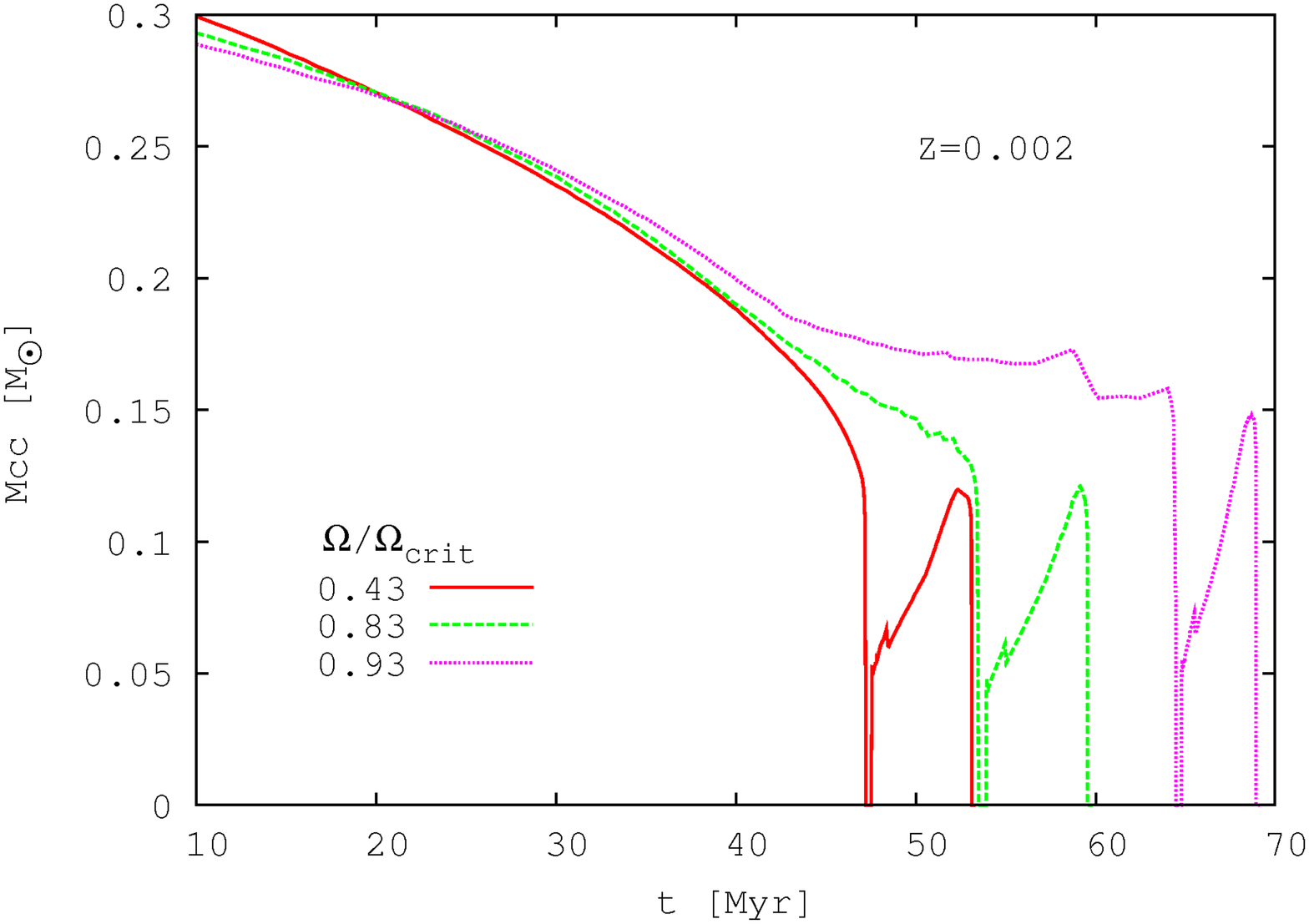}}
\caption{HRD (top) and evolution of the mass of the convective core (bottom) of 7 M$_{\sun}$ models with metallicities of Z=0.014 (left), Z=0.006 (center), and Z=0.002 (right), 
for different rotational rates at the ZAMS.}
\label{HRD_7}
\end{figure*}

As shown by \citet{Meynet2000}, \citet{Ekstrom2008b}, and \citet{Georgy2013a}, when they are governed by the centrifugal
force at the ZAMS, rotating models behave like lower mass models, with a
shift of the tracks towards lower L and T$_{\rm eff}$, compared with non-rotating models. This implies that the size of
the convective core on the ZAMS decreases as a function of
the initial rotational velocity. Later in the evolution, the core is fed with fresh hydrogen by the effects of rotational mixing,
slowing down the decrease in the mass of the convective core (M$_{cc}$), and the newly produced helium
is brought to the radiative zone, and accordingly our models evolve on bluer and more luminous tracks.
Figure \ref{HRD_7} shows the evolutionary tracks (top) and the evolution of the M$_{cc}$ (bottom) of 7 M$_{\sun}$ models.
As already shown by \citet{Georgy2013a}, we find an extreme behaviour for our lower metallicity models: 
during the MS of the models with the highest rotational rate at the ZAMS (or in other words the models with the
largest angular momentum content), the core is nourished with fresh hydrogen brought by rotational mixing
in such a way that the star evolves at almost constant M$_{cc}$. Consequently, the MS lifetimes are significantly longer (see Figure \ref{HRD_7}, bottom). 
For the 7 M$_{\sun}$ model with the largest angular momentum we computed at Z=0.002, the mixing of chemical species during the critical rotation phase 
is extremely efficient in the whole star. This increases the mean molecular weight $\mu$ in the external layers of the star, 
which has a well-known effect on its effective temperature and luminosity, shifting it bluewards in the HRD. This bluewards evolution,  
is typically known as chemically homogeneous evolution \citep{Maeder1987,Yoon2005} (Figure \ref{HRD_7}, top right).
\begin{figure}[h]
\centerline{\includegraphics[width=0.35\textwidth, angle=270]{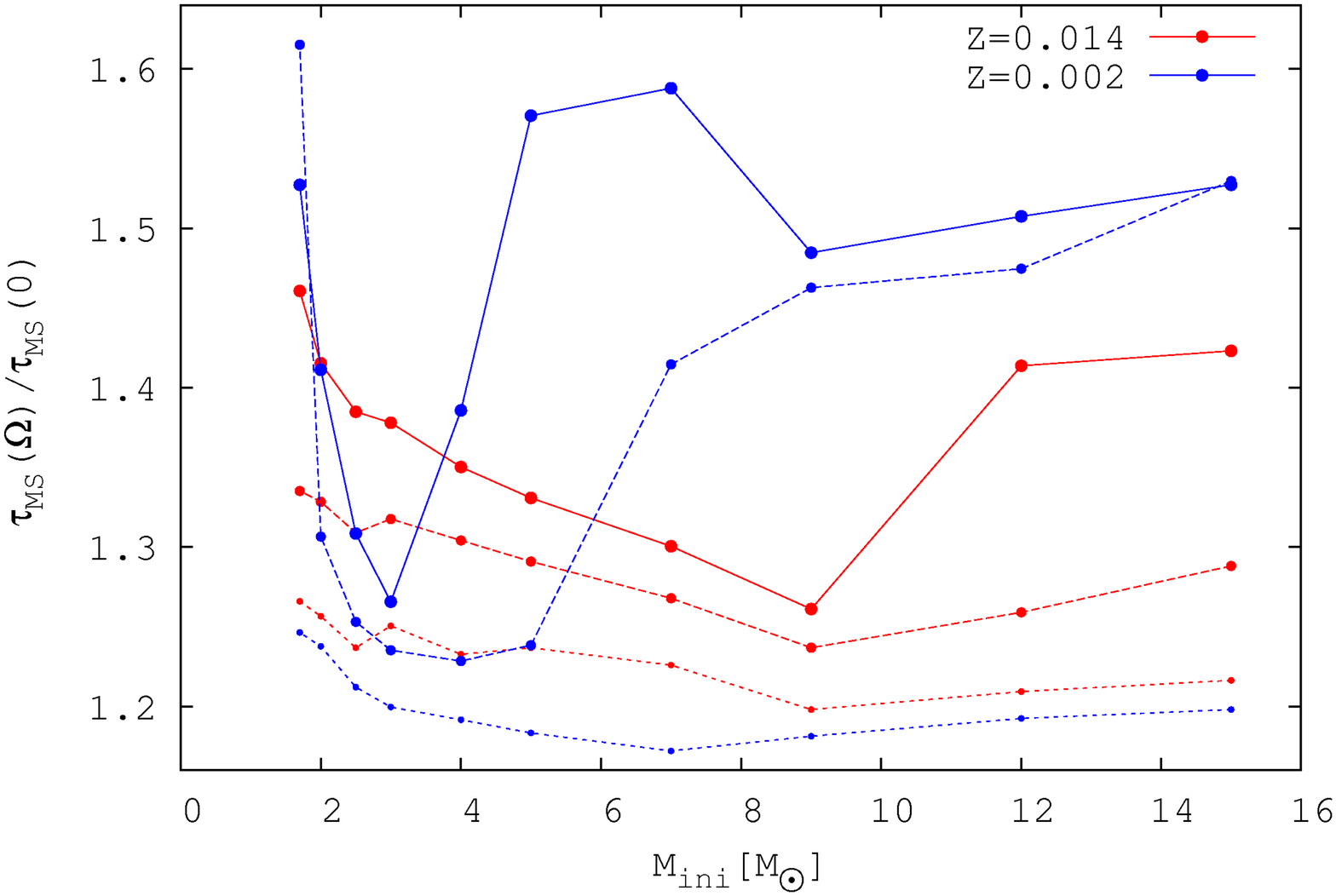}}
\centerline{\includegraphics[width=0.35\textwidth, angle=270]{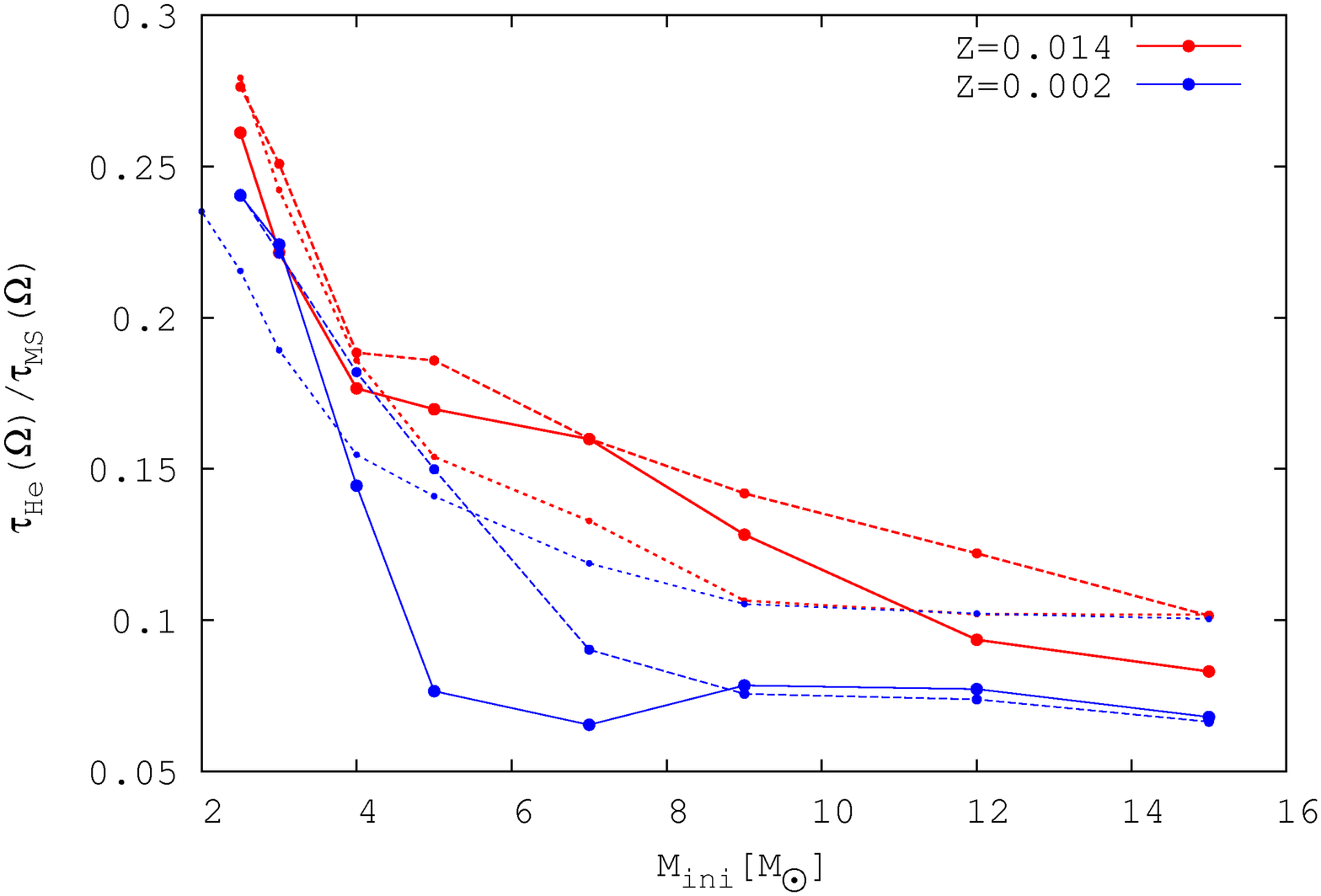}}
\caption{Extension of the MS duration (top) and helium-burning phase (bottom) with respect to the non-rotating case for 
Z=0.014 (red) and Z=0.002 (blue), for clarity we did not plot the results for Z=0.006, which behave much like Z=0.014. 
Three rotational velocity rates are shown: $\sim$0.43\% (dotted line), $\sim$0.83\% (dashed line)
and higher than 90\% of critical (continuous line).}
\label{lifetimesZ}
\end{figure}
The dependence of the MS duration of our new models with mass follows the same trend as was found for lower rotational rates:
because rotational mixing increases the MS lifetimes, at Z=0.014 the
increase amounts to about 15\%-25\% compared with the non-rotating model for stars rotating below 0.5 of the critical limit,
whereas for the most rapid rotators we computed, the variation in the duration of the main sequence can be as
high as 45\%, as seen in Figure \ref{lifetimesZ} (top). 
For lower metallicities, at Z=0.002 the effects of rotation in the duration of the MS are very strong, with an increase of 
more than 25\% for all masses at high rotational rates. Our Z=0.006 models behave much like the solar metallicity case.  
For the smallest and largest masses we computed, this increase of the MS lifetime is higher than 55\%. 
The duration of the helium-burning phase relative to the MS phase decreases with mass and is shorter for Z=0.002. At this metallicity, 
 for masses below 5 M$_{\sun}$ the star burns helium for a significant time, and undergoes a blue loop, which starts when the star is a red supergiant (RSG). 
For higher masses, as remarked by \citet{Georgy2013a}, rapid rotation leads the star to evolve 
directly to the top of the loop, without the first RSG phase. In this case, most of the core He burning occurs in the blue part of the HRD.

\subsection{Surface abundances}

\begin{figure}
\centerline{\includegraphics[width=0.35\textwidth, angle=270]{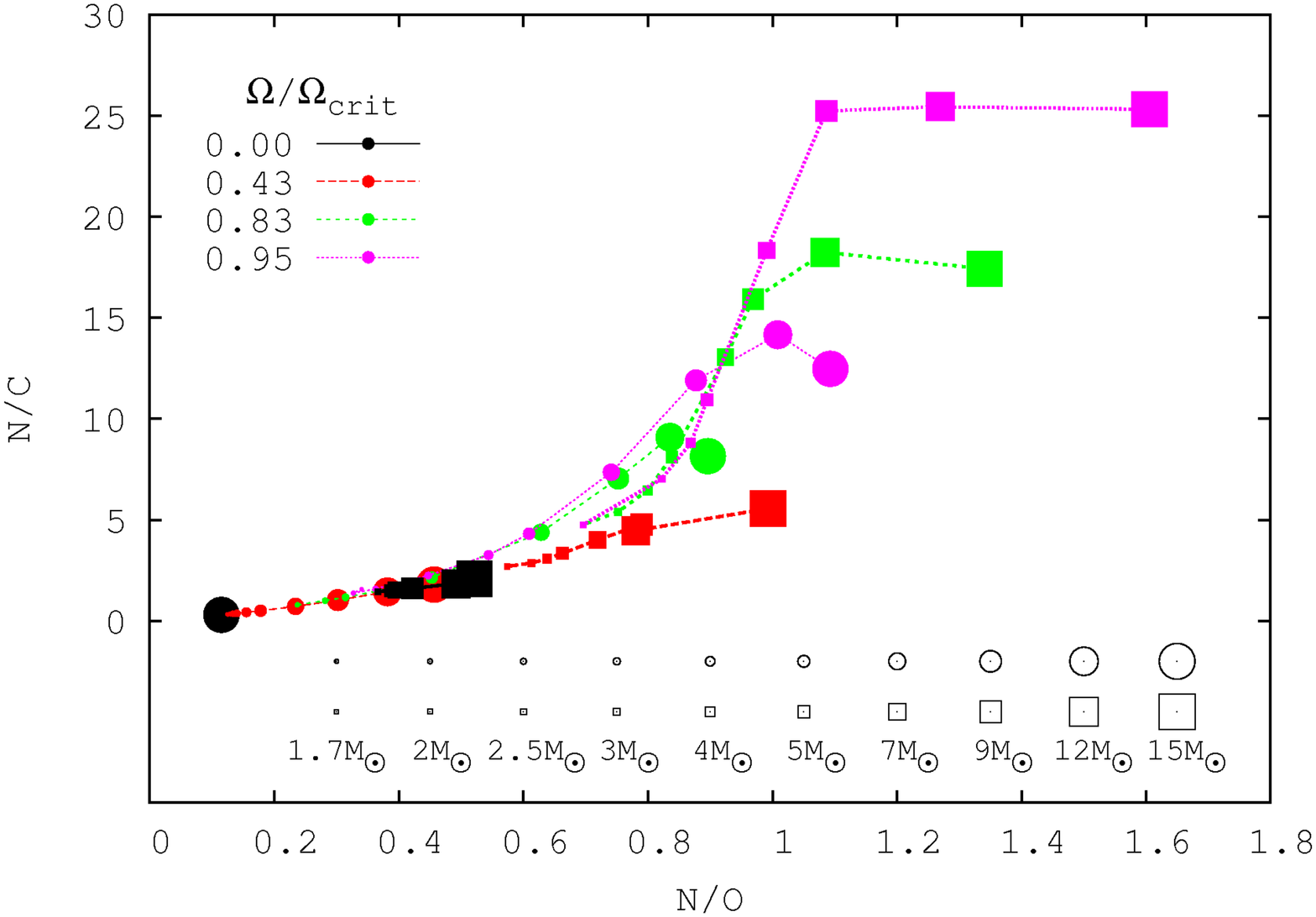}}
\caption{Surface nitrogen to carbon mass ratio versus the nitrogen to oxygen ratio at the end of hydrogen burning (circles) and end of helium burning (squares),
for different stellar masses at Z=0.014. As shown at the bottom, different symbol sizes indicate different stellar masses, from 1.7 M$_{\sun}$ up to 15 M$_{\sun}$.}
\label{NCNO}
\end{figure}

\begin{figure}
\centerline{\includegraphics[width=0.35\textwidth, angle=270]{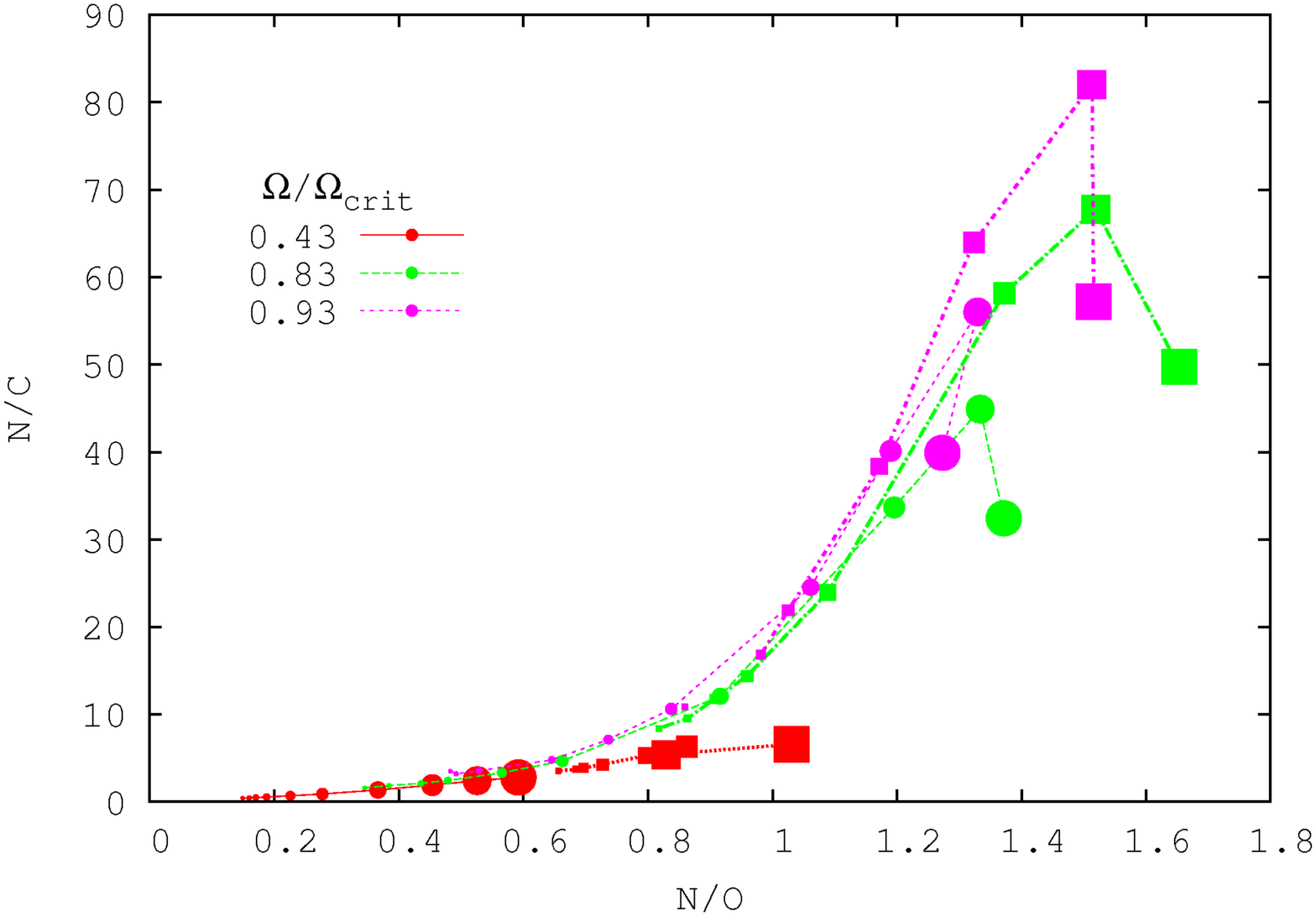}}
\caption{Same as Figure \ref{NCNO}, but with Z=0.006. The non-rotating case is not shown in this plot.}
\label{NCNO_Z006}
\end{figure}

\begin{figure}
\centerline{\includegraphics[width=0.35\textwidth, angle=270]{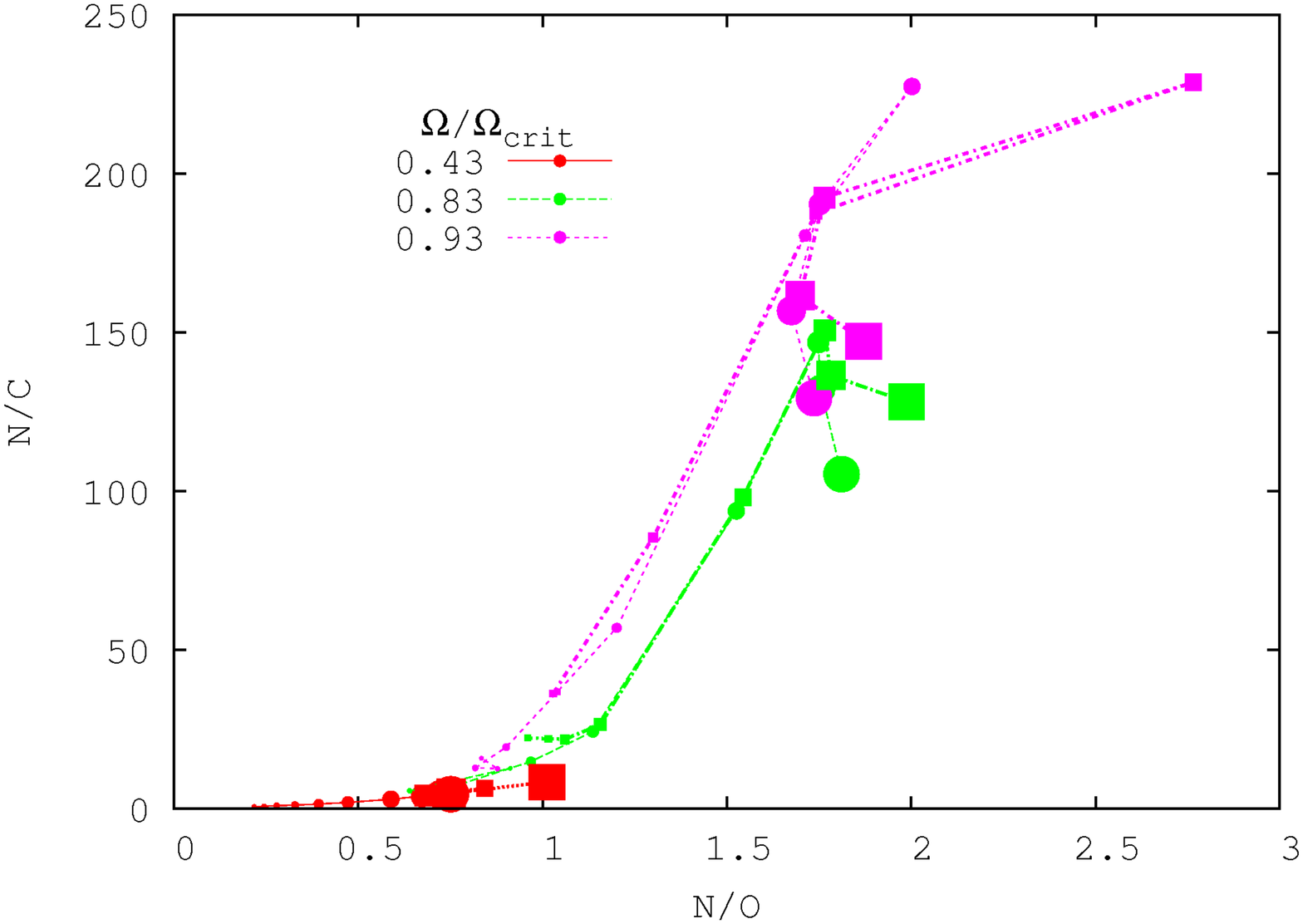}}
\caption{Same as Figure \ref{NCNO}, but with Z=0.002. The non-rotating case is not shown in this plot.}
\label{NCNO_Z002}
\end{figure}

The changes of surface abundances of CNO reflect the actions of the
dominating burning cycle and the mixing mechanisms that occurr in the star.
 \citet{Przybilla2010} showed that a simple analytical derivation 
from the CNO cycle for intermediate mass stars, when the effects of mixing remain modest, leads to a linear relation between the relative enrichments 
  N/C vs. N/O, with a slope $\sim$ 4 \citep[see also][]{Maeder2014}.
We show in Figs. \ref{NCNO}, \ref{NCNO_Z006}, and \ref{NCNO_Z002} the N/C vs. N/O diagram for Z=0.014, Z=0.006 and Z=0.002, respectively, 
for different initial rotational rates and different stellar masses, at the end of the MS and at the end of the helium-burning phase. 
The slowly rotating models ($\Omega/\Omega_{\rm crit}<0.5$) follow the linear trend described in \citet{Przybilla2010}.
Our new rapidly rotating models (magenta) show strong changes in the surface abundances already at the end of the MS. This is
particularly remarkable for Z=0.002, as was already discussed by \citet{Georgy2013a}.

\begin{figure}
\centerline{\includegraphics[width=0.35\textwidth, angle=270]{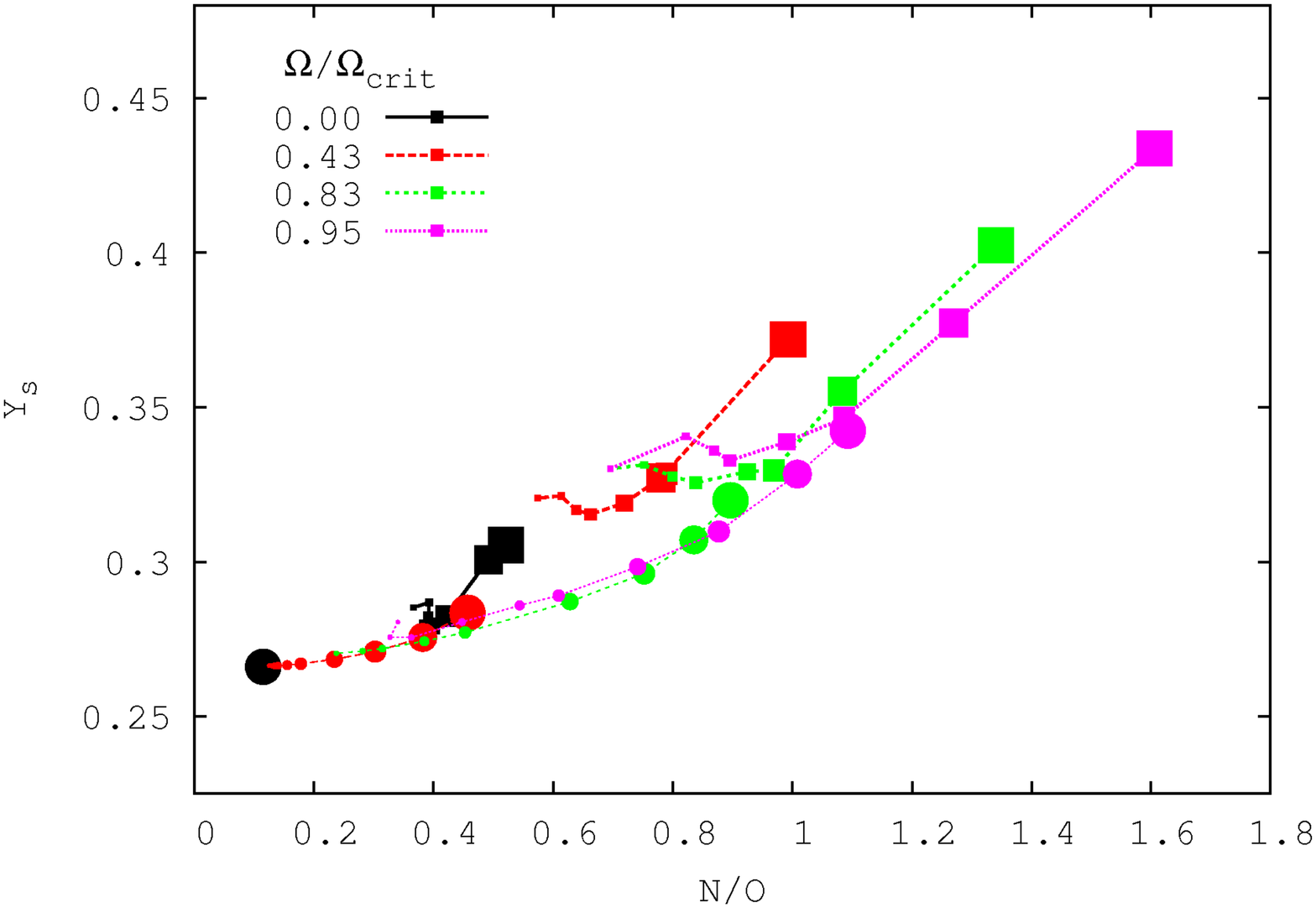}}
\caption{Helium surface abundance (Ys) versus the ratio N/O for stars at the end of the MS (circles) and at the end of the helium burning (squares) 
for models at Z=0.014 with different masses, represented by different point sizes as in Fig. \ref{NCNO}, and different rotational rates at the ZAMS by different colours.} 
\label{surfaceY}
\end{figure}

\begin{figure}
\centerline{\includegraphics[width=0.35\textwidth, angle=270]{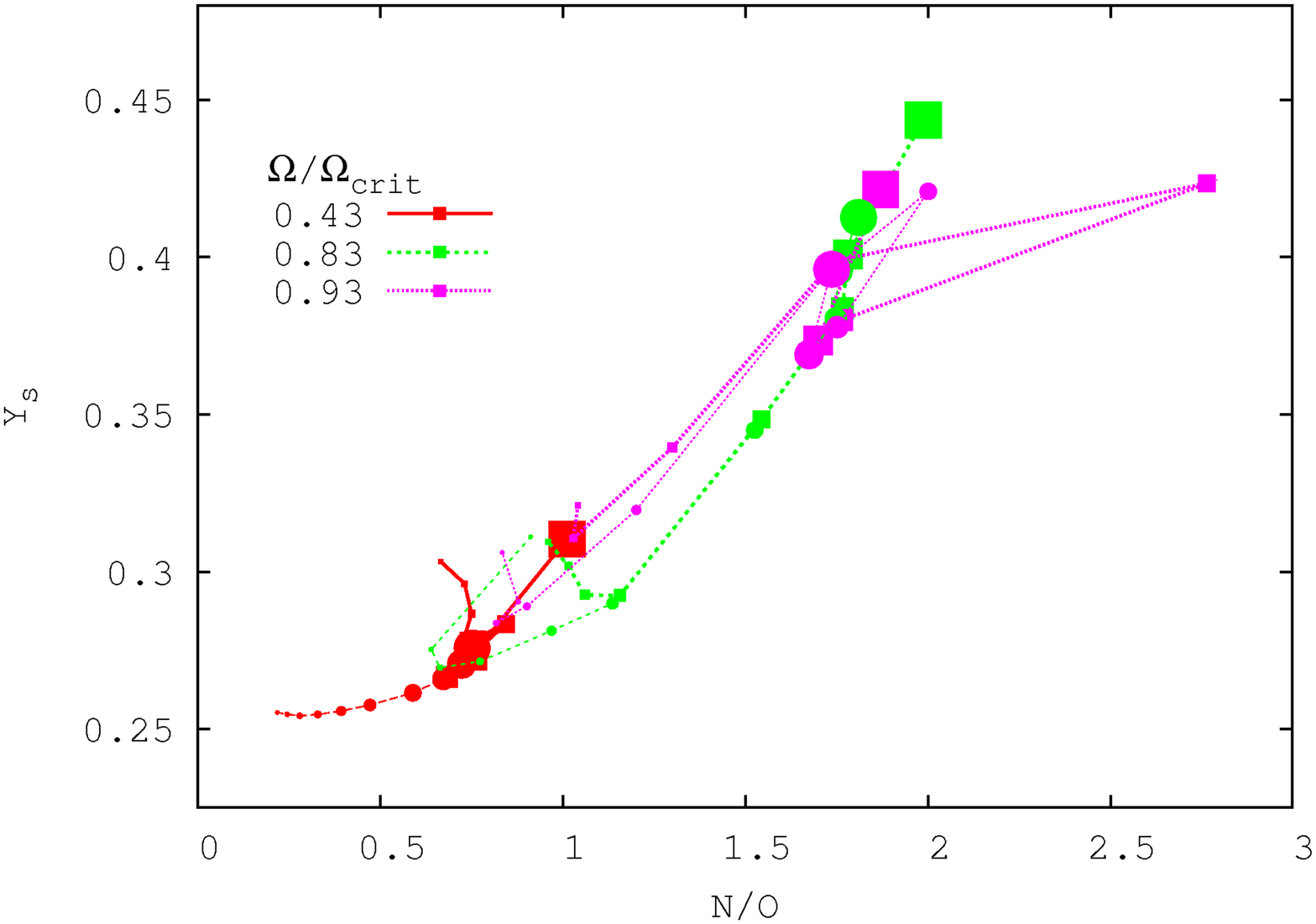}}
\caption{Same as Figure \ref{surfaceY}, but for Z=0.002.}
\label{surfaceY_Z002}
\end{figure}

For the helium surface abundances, \citet{Przybilla2010} showed that MS B-type stars show a roughly constant value ($\sim$ 0.28), while higher enrichments
are obtained for more evolved objects, between 0.3 and 0.4. Figures \ref{surfaceY} and \ref{surfaceY_Z002} show the helium enrichment versus the ratio N/O at the end of the MS
and at the end of the helium burning for Z=0.014 and Z=0.002. At Z=0.014, our slowly rotating models
have a roughly constant helium content at the end of the MS ($\sim$ 0.27-0.3) that is close to the observed values mentioned above, 
in particular for models with masses smaller than 12 M$_{\sun}$. For models with the largest masses 
and rotational rates at the ZAMS the enrichment is somewhat higher, reaching values expected for post-MS objects.   

\begin{figure}
\centerline{\includegraphics[width=0.35\textwidth, angle=270]{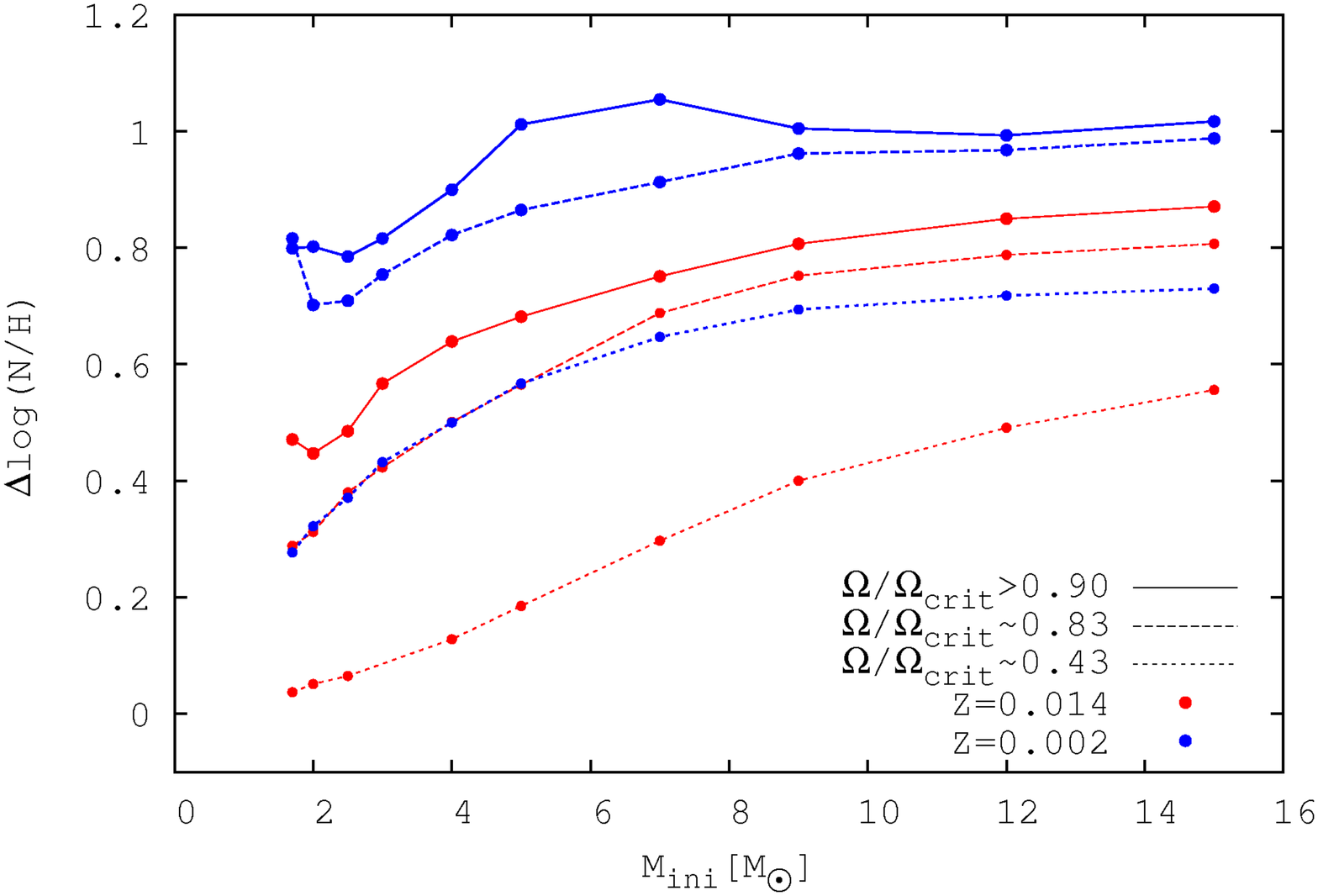}}
\caption{N/H abundance at the end of the main sequence for Z=0.002 (blue) and Z=0.014 (red) and three rotational velocities: $\sim$0.43\% (dotted line),
 $\sim$0.83\% (dashed line), and larger than 90\% of critical (continuous line).}
\label{NHvsM}
\end{figure}

The surface nitrogen enrichments for our rotating models are shown in Figure \ref{NHvsM}. For the metallicities presented here,
the enhancements obtained for the most massive and most rapidly rotating models would be more likely to be detected, in particular towards the end of the MS. 
Even though the chemical enrichments might be detectable in stars, there are very few massive stars that have been critical rotators throughout their whole MS
 to reach the highest values presented in this work. If a MS B-type star rotating close to its critical limit were observed, it would only show a high
 enrichment either if it were very close to their TAMS or if it had been rotating almost critically during most of its lifetime. 
Early B-type stars with large V\,$sin$i that are close to their TAMS are good candidates in which to search for surface enrichments.

\subsection{Evolution of surface velocities and mechanical mass loss}
\begin{figure*}[h]
\centerline{\includegraphics[width=0.32\textwidth, angle=270]{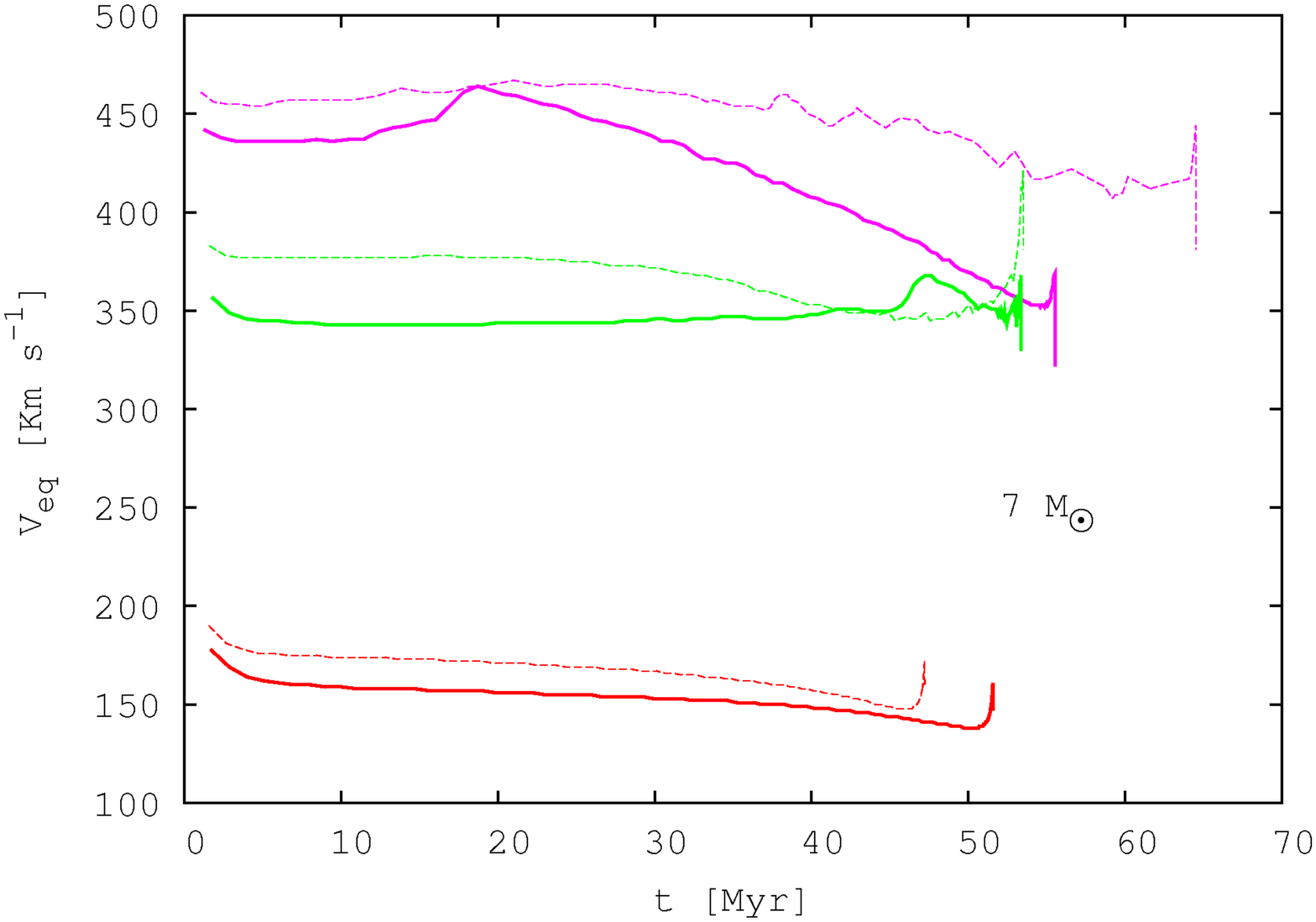}\includegraphics[width=0.32\textwidth, angle=270]{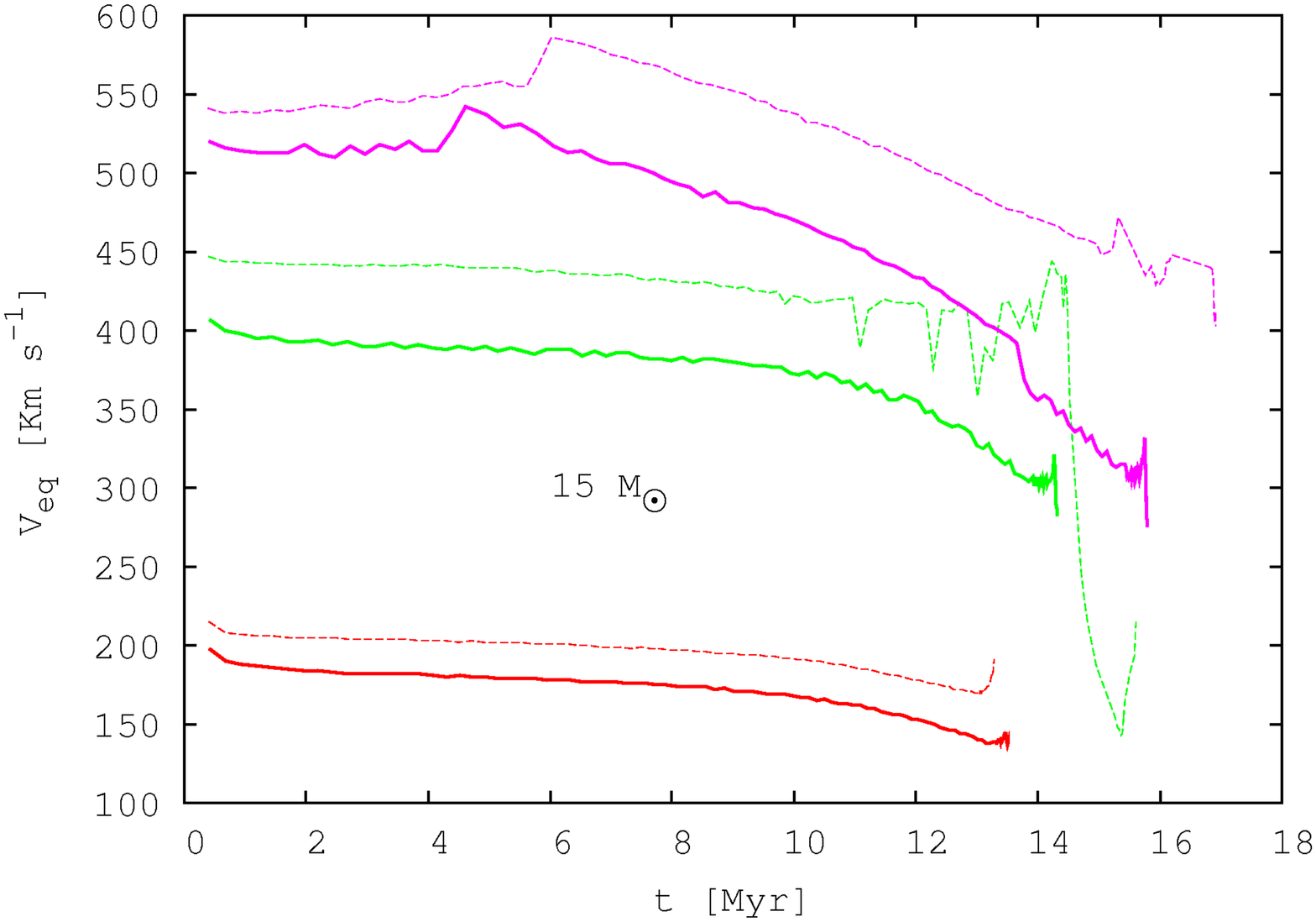}}
\caption{Evolution of the equatorial velocity through the MS phase for 7 M$_{\sun}$ and 15 M$_{\sun}$ models. Continuous lines correspond to Z=0.014, whereas dashed lines correspond to
Z=0.002. Different colours indicate different values of $\Omega/\Omega_{\rm crit}$ at the ZAMS: red corresponds to $0.43$, green to $0.83$, and magenta to our new models
with a large angular momentum content.}
\label{Veq_7}
\end{figure*}

All our new rapidly rotating models with Z=0.002, Z=0.006, and Z=0.014 with masses smaller than 15 M$_{\sun}$
 reach the critical limit soon after the ZAMS, and therefore they lose mass mechanically during a longer time of the MS than the models presented by \citet{Georgy2013a}, 
and the duration of the MS phase is prolonged. The most massive stars, reach the critical limit, but soon stellar winds become efficient 
in removing the excess of angular momentum at the stallar surface, preventing the star from reaching the critical limit.
Table \ref{TabBeModels} shows the duration of the critical rotation phase, the central hydrogen abundance when critical rotation is attained, the total mass lost for these models,
the mass lost via stellar winds and mechanically, and also the mean mechanical mass-loss rate during the critical phase for our new computations. In all the cases, for each mass and metallicity the mean 
mass-loss rates throughout the critical rotation phase remain very similar to the rates we had obtained for models presented in \citet{Granada2013a} with the largest angular momentum content.
We exemplify in Figure \ref{Veq_7} the evolution of the equatorial velocity through the MS phase for 7 M$_{\sun}$ and 15 M$_{\sun}$ models 
for different rotational rates at the ZAMS and for two different metallicities Z=0.002 and Z=0.014. 
We plot our new models that rotate significantly faster throughout the MS than our previous computations with the highest rotational rate at the ZAMS.
The extension in the duration of the MS lifetime is quite remarkable as well, in particular at Z=0.002: at this metallicity, the duration of the MS 
for the models with a large angular momentum content is even longer than at Z=0.014. For the model with 7M$_{\sun}$,  
the time spent on the MS is close to 20\% longer.

\subsection{Metallicity dependency of the surface velocity at the ZAMS}
\begin{figure*}
\centerline{\includegraphics[width=0.64\textwidth, angle=270]{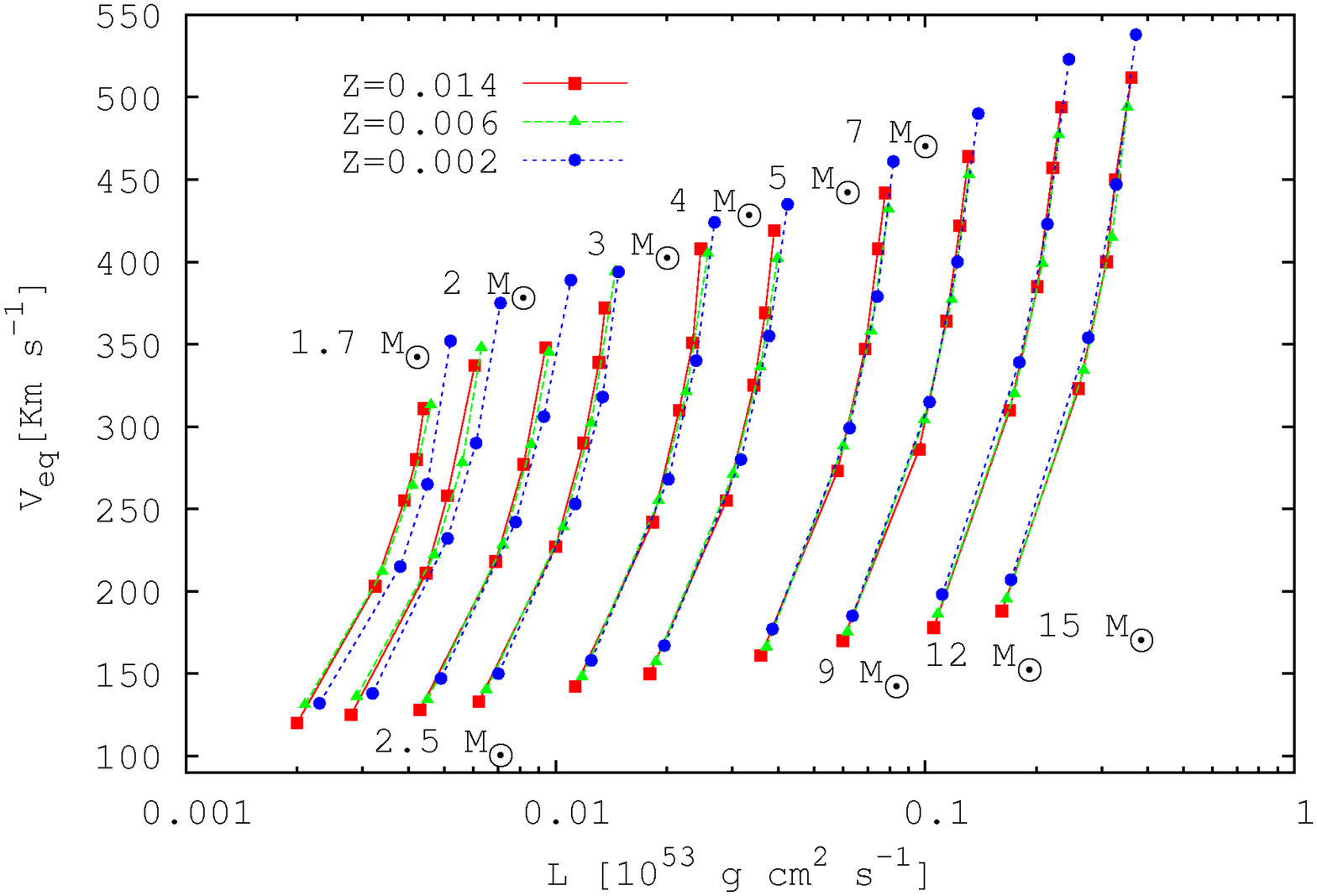}}
\caption{Equatorial velocity at the ZAMS versus the angular momentum content for different masses and metallicities, red squares correspond to Z=0.014, green triangles
to Z=0.006, and blue circles correspond to Z=0.002.}
\label{Ang_Mom}
\end{figure*}

Figure \ref{Ang_Mom} shows the initial angular momentum content versus the rotational velocity at the ZAMS for the models presented in 
Tables \ref{TabListModelsZ014} and \ref{TabListModelsZ002}. We see that 
{\it the equatorial rotational velocity at the ZAMS is almost metallicity independent, 
for the same mass and the same angular momentum content.} 
This is a counterintuitive finding. Because of the differences in opacity, a star at the ZAMS at Z=0.002 is more compact than a star of identical mass at Z=0.014. 
We would expect the more compact star to have a smaller momentum of inertia, and therefore, a faster rotation 
for a given angular momentum content.
We explain below that this is not the case, according to our models.  

The total angular momentum of the star is given by
\begin{equation}
J=\int_\star\rho\,j\,d^3x={8\over3}\pi\int_0^R\rho\,r^4\,\Omega\,dr,\end{equation}

where $j$ and $\rho$ are the local specific angular momentum and density. Calling $s=r/R$ and $\bar{\rho}=M/({4\over3}\pi R^3)$, it follows
\begin{equation}
J=2MR\,{\rm v_{surf}}\int_0^1{\rho\over\bar{\rho}}\,{\Omega\over\Omega_{\rm surf}}\,s^4\,ds.
\label{eq-jv-j}\end{equation}
Eq.~(\ref{eq-jv-j}) shows that for two stars with the same $M$ and $J$, their radii and surface velocities are related by
\begin{equation}
R_1\times{\rm v_{surf\,1}}=R_2\times\rm v_{surf\,2},
\label{eq-jv-rv}\end{equation}

if the relative distribution of the mass (i.e. $\rho/\bar{\rho}$ as a function of $s$)
and of the internal angular velocity profile ($\Omega/\Omega_{\rm surf}$ as a function of $s$)
inside the star are the same.

Observations show that stars in the SMC rotate on average $\sim20-30\%$ faster (in $\rm v_{surf}$), 
than stars in the Milky Way for the mass range considered here \citep[e.g.][]{Martayan2007}.
These observations, together with the compactness of stars at lower metallicities, is normally interpreted as 
satisfying Eq.~(\ref{eq-jv-rv}), giving a ratio
\begin{equation}
{{\rm v_{surf\,02}}\over{\rm v_{surf\,14}}}={R_{14}\over R_{02}}\simeq1.2-1.3 ,
\end{equation}
where $\rm v_{surf}$ indicates the equatorial surface velocity and the subscript $02$ and $14$ indicate SMC and solar metallicities, respectively, 
 in agreement with the observations.
As explained above, this relation implicitely assumes that the relative distribution of the mass and of the angular velocity inside stars of different metallicities 
are the same.

Since we obtain from our models at the ZAMS that  ${\rm v_{surf\,02}}\simeq{\rm v_{surf\,14}}$ and $R_{14}\neq R_{02}$,
we expect that the relative distribution of the mass and/or of the angular velocity must differ between both metallicities. 
We briefly show here that what differs between models of different metallicities is the mass distribution. 

We compare two models with $Z=0.014$ and $Z=0.002$, with the same mass and angular momentum content at the ZAMS, $M_{\rm ZAMS}=9\,M_{\sun}$ and
$J_{\rm ZAMS}=6.81\times10^{51}\rm\,g\,cm^2\,s^{-1}$. In these two cases the internal rotation profile is differential.
For these given quantities, the stellar radii and equatorial velocities are for the larger metallicity, 
$R_{14}=3.71\,R_\sun$ and ${\rm v_{surf\,14}}=198\rm\ km\,s^{-1}=35\%\rm\,v_{crit\,14}$, while for Z=0.002 they are
$R_{02}=3.02\,R_\sun$ and ${\rm v_{surf\,02}}=203\rm\ km\,s^{-1}=33\%\rm\,v_{crit\,02}.$
The ratios presented above are ${R_{14}\over R_{02}}=1.23$ and ${\rm v_{surf\,02}\over v_{surf\,14}}=1.03$, and 
therefore ${R_{14}\over R_{02}}\neq{\rm v_{surf\,02}\over v_{surf\,14}}$.

We compare this result with the case of solid-body rotation,  $\Omega^{\rm sol}=\Omega_{\rm surf}^{\rm sol}$, with identical characteristics as above ($M$, $J$, density profiles and radii).
In this case Eq.~(\ref{eq-jv-j}) reads
\begin{equation}
 J=2MR\,{\rm v_{surf}^{\rm sol}}\int_0^1{\rho\over\bar{\rho}}\,s^4\,ds.
\label{eq-jv-sol}
\end{equation}
Replacing this with the known quantities and integrating Eq.~(\ref{eq-jv-sol}) for the two metallicities, we obtain:
\begin{equation}
{R_{14}\,{\rm v_{surf\,14}^{\rm sol}}\over R_{02}\,{\rm v_{surf\,02}^{\rm sol}}}
={\int_0^1{\rho_{02}\over\bar{\rho}_{02}}\,s^4\,ds\over\int_0^1{\rho_{14}\over\bar{\rho}_{14}}\,s^4\,ds}
={0.040\over0.034}=1.18.
\label{eq-jv-rap2}\end{equation}

Because we know that ${R_{14}\over R_{02}}=1.23$, the velocity ratio is ${\rm v_{surf\,02}^{\rm sol}\over v_{surf\,14}^{\rm sol}}=1.03$, the same as we found 
for a differential rotation profile.

This result shows that the difference between ${R_{14}\over R_{02}}$ and ${\rm v_{surf\,02}\over v_{surf\,14}}$ is fully explained by a difference in the distribution of mass inside the star.

A complete study of this effect and its causes for a wider range in mass and metallicity 
is postponed to a forthcoming paper (Haemmerl\'{e} et al. in prep.).
We recall again that in the context of rotating stellar populations environments with lower metallicity than solar, such as that of the SMC, are claimed to give
rise to stars that rotate faster \citep[e.g.][]{Keller2004,Martayan2007}.   
 If higher rotational velocities were actually more likely to occur at the ZAMS at Z=0.002 than at Z=0.014, and our single evolutionary models were adequate 
to describe the characteristics of rapidly rotating stars, 
this would mean that low-metallicity environments form stars with a larger angular momentum content.
\subsection{Impact on the expected number of rapidly rotating B-type stars with time.}

We have shown throughout this work that rotating models computed assuming solid rotation at the ZAMS adequately represent models with identical metallicity, mass, and 
angular momentum content, but with a differentially rotating internal profile, except for very early in the MS. 
Our pre-MS computations showed that the internal rotational profile has reached a quasi-stationary profile already at the ZAMS.   
This is the same internal rotational profile obtained 
after the sharp decrease in the surface velocity of our models with a flat profile at the ZAMS. Therefore, we propose to 
use the grids presented by \citet{Georgy2013a} (assuming that 
the rotational rate at the ZAMS is the rate when the quasi-equilibrium has been reached)
together with our new models of very rapidly rotating stars throughout the MS, 
to study how rotating stellar populations evolve in time, as we have already done in \citet{Granada2013a}, by using the new Geneva population synthesis code 
(SYCLIST, for synthetic clusters, isochrones, and stellar tracks) presented by \citet{Georgy2014a}. 

We studied how the fraction of stars with surface velocities above a given limit
varies in clusters of various ages. 
We briefly recall that the populations are computed 
in the mass domain of 1.7 - 15 M$_{\sun}$ with initial rotational rates between 0 and 1.
We discretized the mass domain into 1000 mass intervals and the velocity domain into 100 intervals.

Each mass and velocity cell was assigned with the expected number of stars and then normalised 
according to an initial mass function \citep{Salpeter1955}
 and to an initial rotation-rate distribution \citep{Huang2010a}. 
 We then built an evolutionary track for these initial conditions and interpolated between our computed tracks. At each timestep, we checked the
spectral type of the star in each of the cells, and its current
rotation rate. If the star was in the B-type domain\footnote{We define a B star as an object with effective temperature between 10000\,K and 30000\,K.}
and the rotation rate was higher than a given constant (0.70, 0.90, or 0.98), we added the number of stars in the
cell to the total number of stars of a given spectral type that rotates more rapidly than the given constant. The population was
normalised with respect to the total number of B stars at the same time.

Even though the distribution by \citet{Huang2010a} was obtained for Z=0.014, we used it as well to produce a population at Z=0.002 
to compare this with our previous results and study the effects of metallicity alone. An identical initial distribution of rotational rates for both metallicities
corresponds to a distribution of higher rotational velocities at the ZAMS for Z=0.002. This corresponds to a larger 
angular momentum content for objects at lowe metallicities, which is in principle coherent with observations of rotating B-type stars. 

\begin{figure}[h]
\centerline{\includegraphics[width=0.5\textwidth, angle=270]{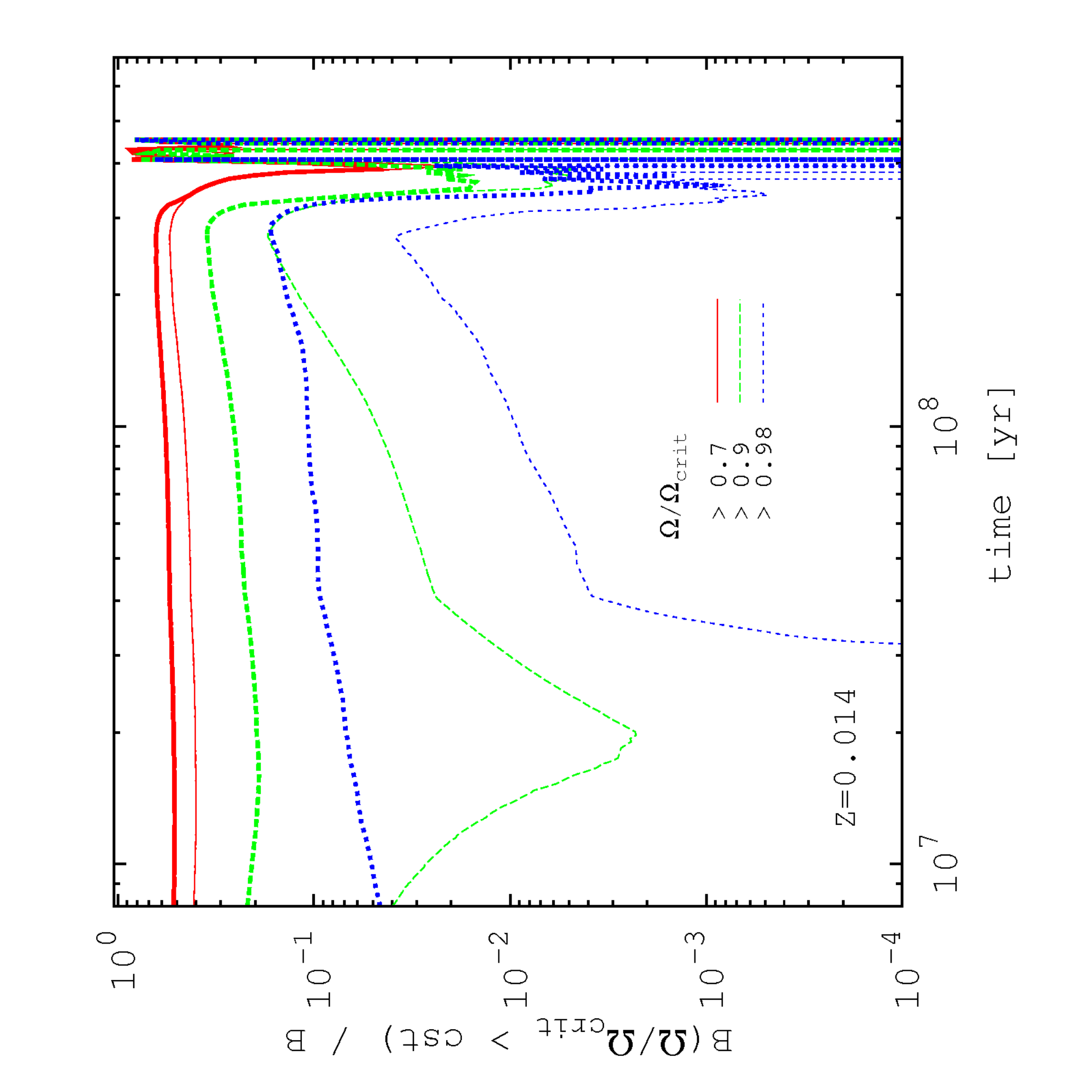}}
\caption{Time evolution of the fraction of B-type stars that rotate faster than a certain $\Omega/\Omega_{crit}$ for Z=0.014. Red continuous lines correspond 
to the fraction of stars rotating with $\Omega/\Omega_{crit}$>0.7, green dashed lines to those with $\Omega/\Omega_{crit}$>0.9, and blue dotted lines to $\Omega/\Omega_{crit}$>0.98.
Thin lines correspond to the ZAMS rotational rates for solidly rotating models shown in \citet{Granada2013a}, whereas the thicker lines
 are our new results obtained assuming that the rotational rate at the ZAMS is the rate when the quasi-equilibrium has been reached.}
\label{PopEvolZ014}
\end{figure}

\begin{figure}[h]
\centerline{\includegraphics[width=0.5\textwidth, angle=270]{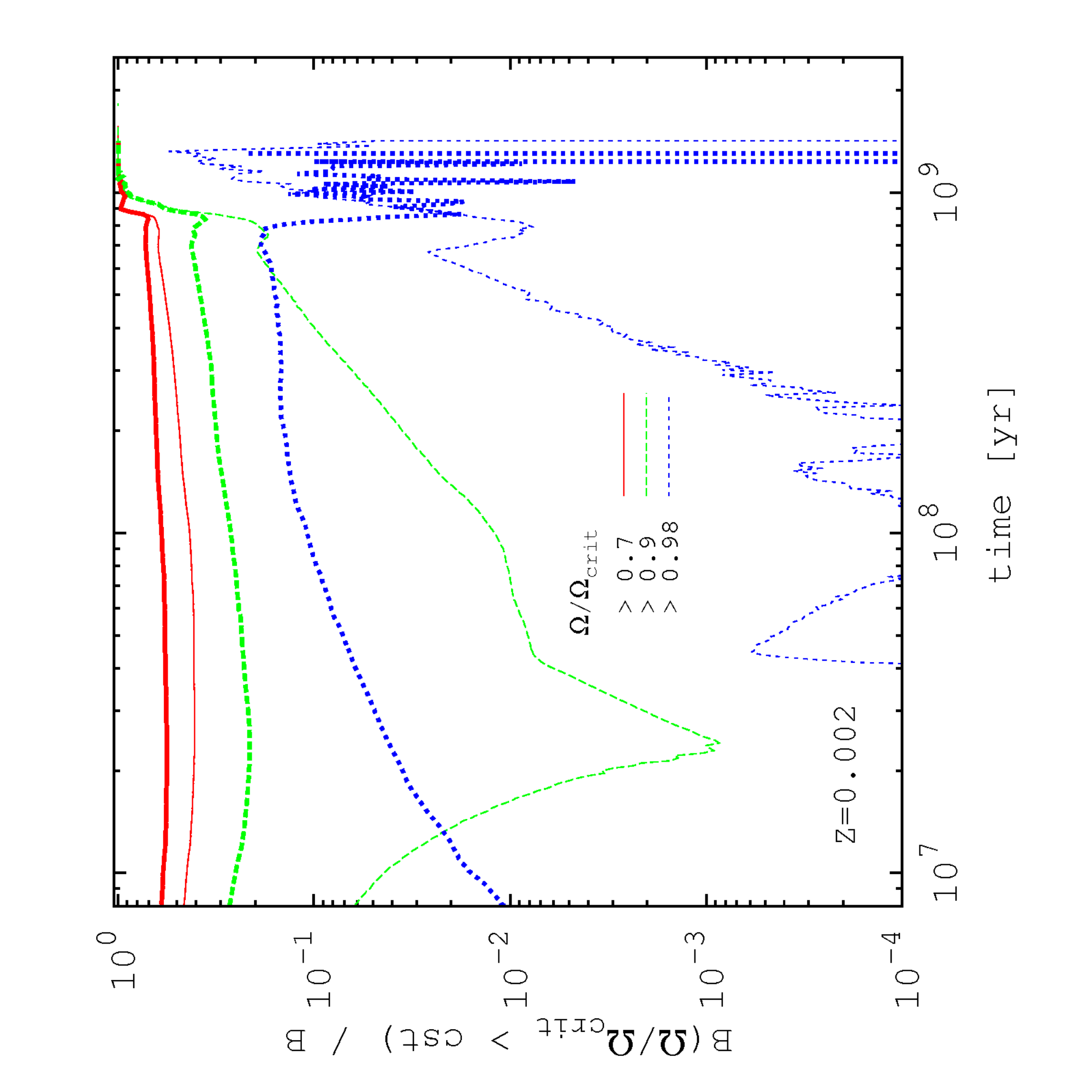}}
\caption{Idem. figure \ref{PopEvolZ014} but for Z=0.002.}
\label{PopEvolZ002}
\end{figure}

Figures \ref{PopEvolZ014} and \ref{PopEvolZ002} show the time evolution of stars rotating faster than 0.70, 0.90, and 0.98 in terms of $\Omega/\Omega_{crit}$ (these values correspond to
0.51, 0.73, and 0.88 in terms of $v/v_{crit}$, respectively) for Z=0.002 and Z=0.014.


As expected, our new results lead to higher percentages of rapidly rotating stars. The effect is moderate for the fraction of stars rotating
with $\Omega/\Omega_{crit}$>0.7, whereas the changes are strong for the group of the most rapidly rotating stars. We find now that there is always a non-zero
probability of having stars rotating with $\Omega/\Omega_{crit}$>0.98 at all ages, which can become higher than 10\% of B stars towards the TAMS of late B-type 
stars, where the existing B stars are rapidly rotating objects. 

\section{Conclusions}

We have shown that our grids of rotating models computed
assuming solid rotation at the ZAMS \citep{Georgy2013a} adequately represent models
with a different initial internal rotation profile, but identical metallicity, mass, and angular momentum content, except for very
early in the MS. We propose to use these grids in the future to build synthetic populations, assuming
 as the initial rotational velocity the velocity obtained when the star has reached the quasi-equilibrium state described in Section \ref{initialZ}.

By using differentially rotating stellar models at the ZAMS obtained from pre-MS calculations, we were able to produce 
stellar evolution models with larger angular momentum 
than those presented in our previous work for three different metallicities and masses between 1.7 and 15 M$_{\sun}$. 
Objects that are rotating close to the critical rotational rate throughout the whole MS have a more extreme
behaviour than those that reach the critical limit later in the MS. This is particularly important in low-metallicity environments, 
in which the stellar lifetimes, the evolutionary tracks in the HRD, and the surface 
enrichment are strongly modified by the effects of rapid rotation throughout the MS.

Interestingly and counterintuitively, we found that for a certain mass and angular momentum content, 
the equatorial velocity at the ZAMS depends very weakly on the metallicity. This is basically due to differences in the internal mass distribution. 
We will discuss this result thoroughly in a forthcoming paper (Haemmerl\'e et al., in preparation).

We count now on models with a large angular momentum content that rotate close to the critical limit throughout the whole MS.
Our next step will be to use these new models together with the grids by \citet{Georgy2013a} to compare synthetic rotating stellar populations with SYCLIST, 
obtained with the new Geneva population synthesis code \citep{Georgy2014a}, with observed rotating populations.
\begin{acknowledgements}
We thank Georges Meynet, Cyril Georgy, and the anonymous referee of this manuscript for their valuable comments.
\end{acknowledgements}
\bibliographystyle{aa}
\bibliography{citas}
\end{document}